\documentclass[12pt, letterpaper]{article}
\usepackage[utf8]{inputenc}

\usepackage{forest}
\usepackage[ruled,vlined]{algorithm2e}
\usepackage{mathtools}
\usepackage{graphicx}
\usepackage{multirow}
\usepackage{amssymb} 
\usepackage{amsthm}
\usepackage{amsmath}
\usepackage{fullpage}
\usepackage[longnamesfirst]{natbib}
\usepackage{enumerate}
\usepackage{amsfonts, bm, amsmath, color, natbib, rotating, amsthm, multicol, epstopdf, verbatim, hyperref}

\usepackage{authblk}
\usepackage{soul}
\usepackage{xr}
\bibpunct{(}{)}{;}{a}{,}{,}

\usepackage{setspace}
\theoremstyle{plain} \newtheorem{theorem}{Theorem}  \newtheorem{lemma}{Lemma} \newtheorem{corollary}{Corollary}

\theoremstyle{definition} \newtheorem{definition}{Definition}  \newtheorem{example}{Example}

\theoremstyle{remark} \newtheorem*{remark}{Remark}  
\begin{document}

\newif\ifblinded

\ifblinded
\title{Fast, Optimal, and Targeted Predictions \\ using Parametrized Decision Analysis}
\author{\vspace{-15mm}}
\else

\title{\Large Fast, Optimal, and Targeted Predictions \\ using Parametrized Decision Analysis}
\author{\large \vspace{-3mm} Daniel R. Kowal\thanks{Dobelman Family Assistant Professor, Department of Statistics, Rice University (\href{mailto:Daniel.Kowal@rice.edu}{daniel.kowal@rice.edu}).}} 

\fi

\date{}

\maketitle
  
  \vspace{-15mm}
\begin{abstract}
Prediction is critical for decision-making under uncertainty and lends validity to statistical inference. With \emph{targeted prediction}, the goal is to optimize predictions for specific decision tasks of interest, which we represent via functionals. Although classical decision analysis extracts predictions from a Bayesian model, these predictions are often difficult to interpret and slow to compute. Instead, we design a class of \emph{parametrized actions} for Bayesian decision analysis that produce optimal, scalable, and simple targeted predictions. For a wide variety of action parametrizations and  loss functions--including linear actions with sparsity constraints for targeted variable selection---we derive a convenient representation of the optimal targeted prediction that yields efficient and interpretable solutions. Customized out-of-sample predictive metrics are developed to evaluate and compare among targeted predictors. Through careful use of the posterior predictive distribution, we introduce a procedure that identifies a set of near-optimal, or \emph{acceptable} targeted predictors, which provide unique insights into the features and level of complexity needed for accurate targeted prediction. Simulations demonstrate excellent prediction, estimation, and variable selection capabilities. Targeted predictions are constructed for physical activity data from the National Health and Nutrition Examination Survey (NHANES) to better predict and understand the characteristics of intraday physical activity. 
\end{abstract}

\section{Introduction}
Prediction is a cornerstone of statistical analysis: it is essential for decision-making under uncertainty and provides validation for inference \citep{Geisser1993}. Predictive evaluations are crucial for model comparisons and selections \citep{gelfand1992model} and offer diagnostic capabilities for detecting model misspecification \citep{gelman1996posterior}. More subtly, predictions provide an access point for model interpretability: namely, via identification of the model characteristics or variables which matter most for accuracy. However, the demands of many datasets---which can be high-dimensional, high-resolution, and multi-faceted---often necessitate sophisticated and complex models. Even when such models predict well, they can be cumbersome to deploy and difficult to summarize or interpret. 

%




Our focus is \emph{targeted prediction}, where predictions are customized for the decision tasks of interest. The translation of models into actionable decisions requires predictive quantities in the form of \emph{functionals} of future or unobserved data. Predictions should be optimized for these decision tasks---and targeted to the relevant functionals.  The target is fundamental for defining the correct (predictive) likelihood \citep{Bjornstad1990}. Absent specific functionals of interest, targeted prediction offers a path for interpretable statistical learning: the functionals probe the data-generating process to uncover the predictability of distinct attributes. 



\begin{figure}[h!]
\begin{center}
\includegraphics[width=.49\textwidth]{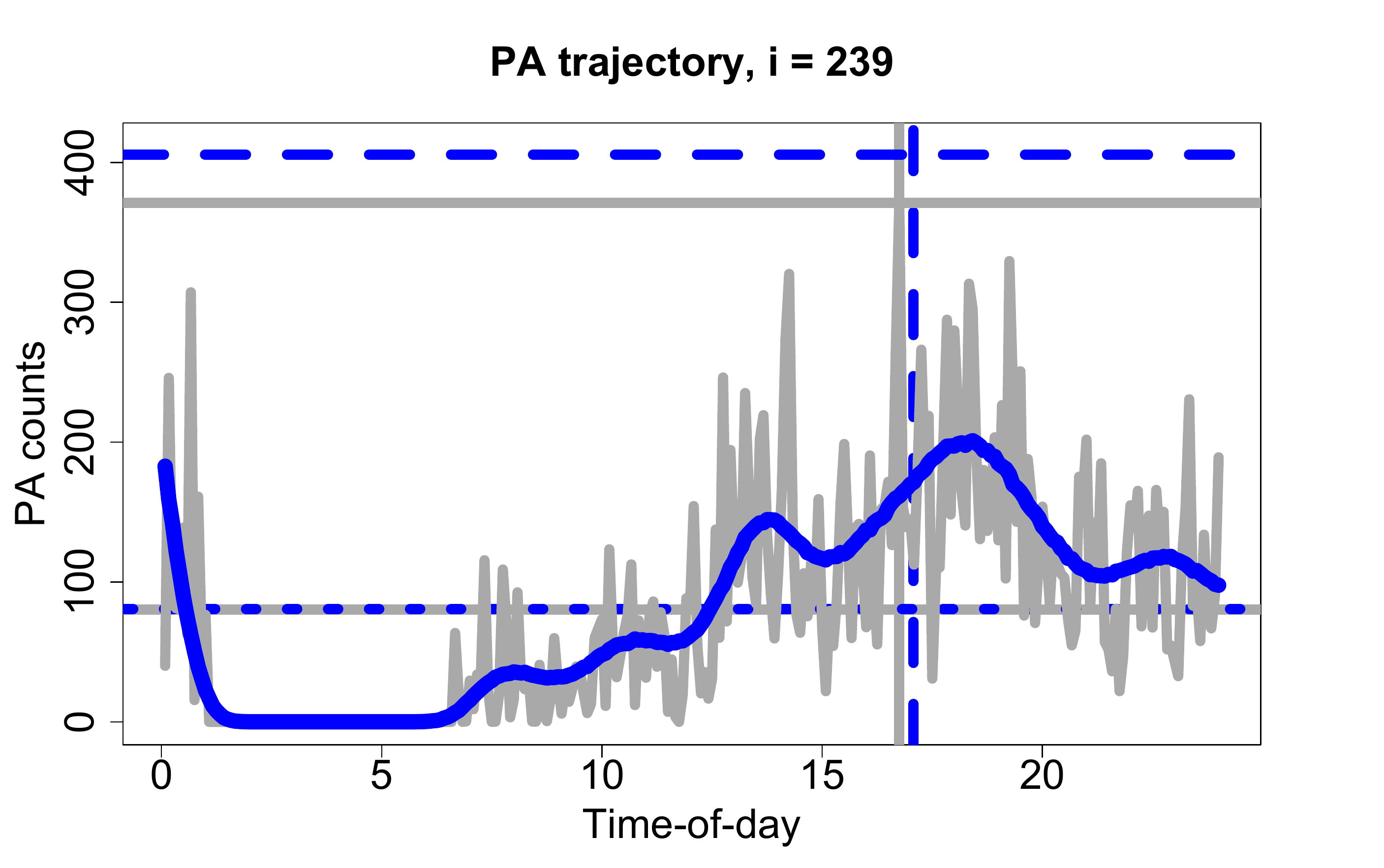}
\includegraphics[width=.49\textwidth]{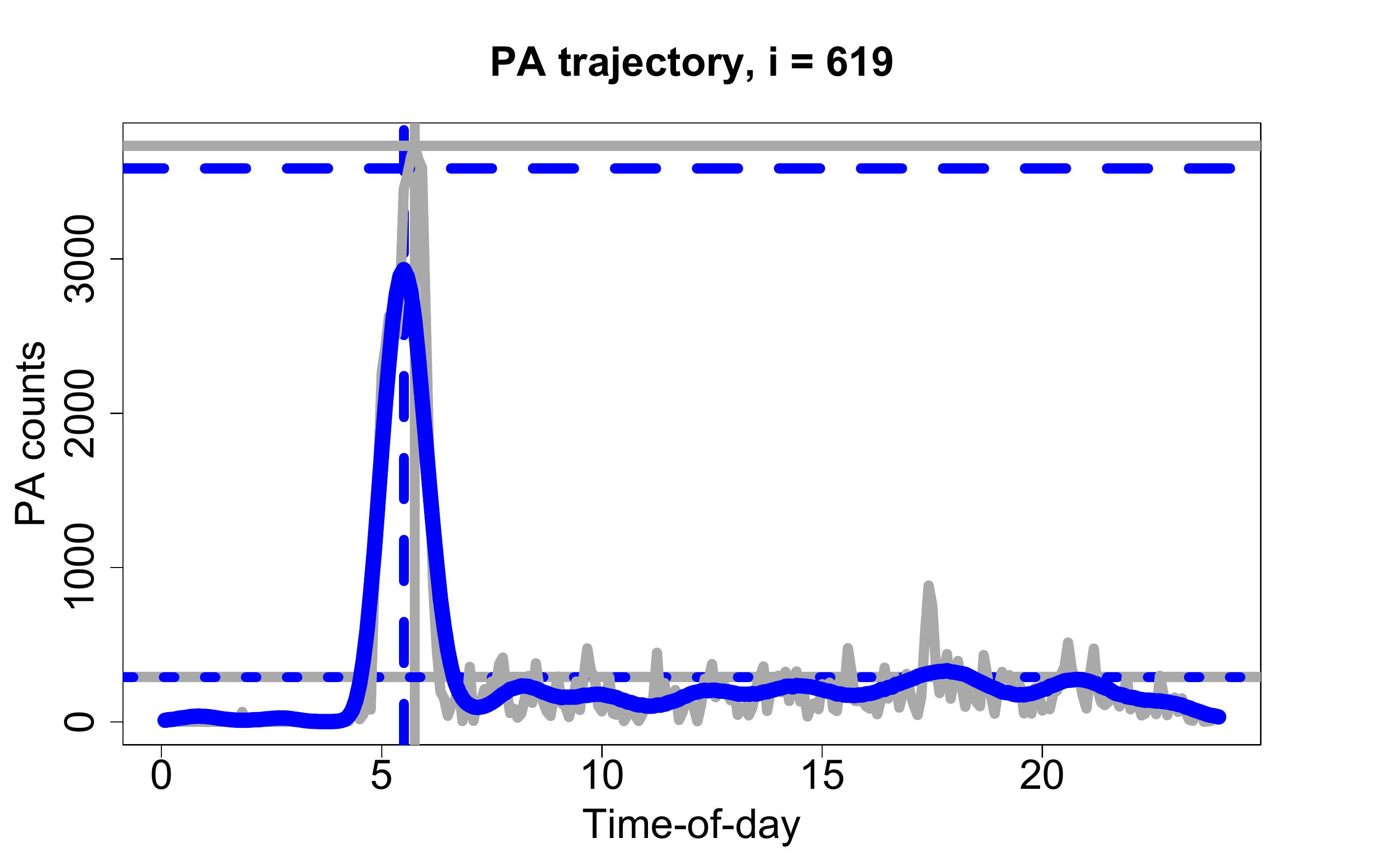}
\caption{\small 
Intraday physical activity (gray line) and fitted values (blue line) for two subjects under model  $\mathcal{M}$ in \eqref{star-round}-\eqref{star-reg}. The lines denote the empirical (solid gray) and predictive expected value (dashed blue) of \texttt{avg} (lower horizontal), \texttt{max} (upper horizontal), and \texttt{argmax} (vertical).
\label{fig:max-pa}}
\end{center}
\end{figure}

To illustrate these points, we display wearable device data from the National Health and Nutrition Examination Survey (NHANES) in Figure~\ref{fig:max-pa}. Physical activity (PA) trajectories are modeled as functional data and accompanied by subject-specific covariates; descriptions of the data and the model are in Section~\ref{pa}. Scientific interest does not reside exclusively with these intraday profiles: we are also interested in functionals of the trajectories. Figure~\ref{fig:max-pa} shows several such functionals: the average activity (\texttt{avg}), the peak activity level (\texttt{max}), and the time of peak activity (\texttt{argmax}).  These functionals summarize daily PA and describe clear sources of variability in PA among the individuals. Other features are discernible, such as sedentary behavior and periods of absolute inactivity, and are investigated in Section~\ref{pa}. However, Bayesian model-based point predictions alone do not explain what drives the variability among individuals and can be slow to compute out-of-sample. 

Our goal is construct targeted predictions that improve accuracy, streamline decision making, and highlight the model attributes and covariates that matter most for prediction---which notably may differ among functionals. Building upon classical decision analysis, we introduce \emph{parametrized actions} that extract optimal, simple, and fast predictions under any Bayesian model $\mathcal{M}$. 
The parameterizations exploit familiar model structures, such as linear, tree, and additive forms, while the actions minimize a posterior predictive expected loss that is customized for each functional. For a broad class of parametrized actions and loss functions, we derive a convenient representation of the optimal targeted prediction that yields efficient and interpretable solutions. These  solutions can be computed using existing software packages for penalized regression, which allows for widespread and immediate deployment of the proposed techniques.  The targeted predictions are constructed simultaneously for multiple functionals based on a single  $\mathcal{M}$, which avoids the need to re-fit a Bayesian model for each functional. While intrinsically useful for  prediction, the elicitation of multiple targeted predictors is also informative for understanding and summarizing the model $\mathcal{M}$ posterior. 


A key feature of our approach is the use of the model $\mathcal{M}$ predictive distribution to provide uncertainty quantification for out-of-sample predictive evaluation. We design a procedure to identify not only the most accurate targeted predictor, but also any predictor that performs nearly as well with some nonnegligible predictive probability. This strategy emerges as a Bayesian representation of the \emph{Rashomon effect}, which observes that there often exists a multitude of acceptably accurate predictors \citep{breiman2001statistical}. The set of \emph{acceptable} predictors is informative: it describes the shared characteristics and level of complexity needed for near-optimal targeted prediction. We do not require any re-fitting of $\mathcal{M}$ 
and instead design an efficient algorithm to approximate the relevant out-of-sample predictive quantities for each functional. The proposed methods are applied to both simulated and real data and demonstrate excellent prediction, estimation, and selection capabilities.

There is a rich literature on the use of decision analysis to extract information from a 
Bayesian model. \cite{Bernardo2009} provide foundational elements, while \cite{vehtari2012survey} give a prediction-centric survey. \cite{maceachern2001decision} and \cite{GutierrezPena2005} use decision analysis to summarize Bayesian nonparametric models.   The proposed methods expand upon a line of research for \emph{posterior summarization}, most commonly for Bayesian variable selection, advocated by \cite{lindley1968choice} and rekindled  by \cite{hahn2015decoupling}. These techniques have been adapted for seeming unrelated regressions \citep{Puelz2017},
graphical models \citep{Bashir2019},
nonlinear regressions \citep{woody2019model},
functional regression \citep{kowal2019bayesianfosr},
and time-varying parameter models \citep{Huber2020}. 
Alternative approaches combine linear variable selection with Kullback-Leibler  distributional approximations \citep{goutis1998model,nott2010bayesian,tran2012predictive,Crawford2019,piironen2018projective}. 
In general, these methods focus on global summarizations of a particular model $\mathcal{M}$ posterior distribution. 
By comparison, our emphasis on \emph{predictive functionals} adds specificity and a direct link to the observables, which provides a solid foundation for (out-of-sample) predictive evaluations and broadens applicability among Bayesian models with different parameterizations.

The remainder of the paper is organized as follows. Section~\ref{target} introduces predictive decision analysis for optimal targeted prediction.  Section~\ref{out-of-sample} develops the methods and algorithms for predictive evaluations and comparisons. A simulation study is in Section~\ref{sims}. The PA data are analyzed in Section~\ref{pa}. Section~\ref{discussion} concludes.   Online supplementary material includes methodological generalizations and further examples, computational details, additional results for the simulated and PA data, proofs, and  \texttt{R} code to reproduce the  analyses.

\section{Targeted point prediction}\label{target}
Consider the paired data $\{\bm x_i, \bm y_i\}_{i=1}^n$ with $p$-dimensional covariates $\bm x_i$ and $m$-dimensional response $\bm y_i$. The response variables $\bm y_i$ may be univariate ($m=1$), multivariate  ($m > 1$), or functional data with $\bm y_i = (y_i(\tau_1,),\ldots, y_i(\tau_m))'$ observed on a domain $\mathcal{T} \subset \mathbb{R}^d$. Suppose we have a satisfactory Bayesian model $\mathcal{M}$ parametrized by $\bm \theta$ with posterior $p_\mathcal{M}(\bm \theta | \bm y)$. The requisite notion of ``satisfactory" is made clear below, but fundamentally $\mathcal{M}$ should encapsulate the modeler's beliefs about the data-generating process and demonstrate empirically the ability to capture the essential features of the data. Although these criteria are standard for Bayesian modeling, they often demand highly complex and computationally intensive models. There is broad interest in extracting simple, accurate, and computationally efficient representations or summaries of $\mathcal{M}$, especially for prediction.

Our approach builds upon Bayesian decision analysis. First, we target the \emph{predictive functionals} $h_1(\bm{\tilde y}), \ldots, h_J(\bm{\tilde y})$, where each $h_j$ is a functional of interest and $\bm{\tilde y} \sim p_\mathcal{M}(\bm{\tilde y} | \bm y)$ is the predictive distribution of unobserved data at covariate value $\bm{\tilde x}$ and conditional on observed data. Each $h_j$ reflects a prediction task: often the data $(\bm x, \bm y)$ are an input to a system $h_j$, which inherits predictive uncertainty when $\bm y$ has not yet been observed. Alternatively, the functionals $\{h_j\}$ can be selected to provide distinct summaries of the model $\mathcal{M}$. Next, we introduce a \emph{parametrized action} $g(\bm{\tilde x}; \bm \delta)$, which is a point prediction of $h(\bm{\tilde y})$ at $\bm{\tilde x}$ with unknown parameters $\bm \delta$. The role of $g$ is to produce interpretable, fast, and accurate predictions targeted to $h$. Examples include linear, tree, and additive forms, but $g$ is not required to match the structure of $\mathcal{M}$. The targeted predictions are not burdened by the complexity required to capture the global distributional features of  $p_\mathcal{M}(\bm \theta | \bm y)$ or $p_\mathcal{M}(\bm{\tilde y} | \bm y)$---which may be mostly irrelevant for predicting any particular $h_j(\bm{\tilde y})$---yet use the full posterior distribution under $\mathcal{M}$ to incorporate all available data. Lastly, we leverage the model $\mathcal{M}$ predictive distribution to quantify and compare the \emph{out-of-sample} predictive accuracy of each parametrized action. Using this information, we assemble a collection of near-optimal, or \emph{acceptable} targeted predictors, which offers unique insights into the predictability of $h_j(\bm{\tilde y})$.

For any functional $h_j = h$, predictive accuracy is measured by a loss function $\mathcal{L}_0 \{h(\bm{\tilde y}), g(\bm{ \tilde x}; \bm{\delta}) \}$, which determines the loss from predicting  $g(\bm{ \tilde x}; \bm{\delta}) $ when $h(\bm{\tilde y})$ is realized. To incorporate multiple covariate values $\mathcal{\widetilde X} \coloneqq \{\bm{\tilde x}_i\}_{i=1}^{\tilde n}$, we introduce an aggregate loss function 
$$\mathcal{\bar L}_0\big[\{h(\bm{\tilde y}_i), g(\bm{ \tilde x}_i; \bm{\delta}) \}_{i=1}^{\tilde n}\big]  \coloneqq {\tilde n}^{-1}\sum_{i=1}^{\tilde n} \mathcal{L}_0 \{h(\bm{\tilde y}_i), g(\bm{ \tilde x}_i; \bm{\delta}) \},$$
 where each $\bm{\tilde y}_i$ is the predictive variable at $\bm{\tilde x_i}$ under model $\mathcal{M}$. The choice of $\mathcal{\widetilde X}$ can be distinct from the original covariates $\{\bm x_i\}_{i=1}^n$, for example to customize predictions for specific designs or subpopulations of interest, yet still leverages the full posterior distribution under model $\mathcal{M}$. We augment the aggregate loss with a complexity penalty $\mathcal{P}$ on the unknown parameters $\bm \delta$: 
$$ \mathcal{\bar L}_\lambda\big[\{h(\bm{\tilde y}_i), g(\bm{ \tilde x}_i; \bm{\delta}) \}_{i=1}^{\tilde n}\big] \coloneqq   \mathcal{\bar L}_0\big[\{h(\bm{\tilde y}_i), g(\bm{ \tilde x}_i; \bm{\delta}) \}_{i=1}^{\tilde n}\big] + \lambda \mathcal{P}(\bm \delta),$$ 
where $\lambda \ge 0$ indexes a path of parameterized actions and determines the tradeoff between predictive accuracy ($ \mathcal{\bar L}_0$) and complexity ($\mathcal{P}$).

 Since  $\mathcal{\bar L}_\lambda$ depends on a random quantities $\{\bm{\tilde y}_i\}_{i=1}^{\tilde n}$, Bayesian decision analysis proceeds by optimizing for $\bm \delta$ over the joint posterior predictive distribution $p_\mathcal{M}(\bm{\tilde y}_1,\ldots, \bm{\tilde y}_{\tilde n} | \bm y)$: 
\begin{equation}\label{action}
\bm{\hat \delta}_{\mathcal{A}}  \coloneqq 
\arg\min_{\bm \delta} \mathbb{E}_{[\bm{\tilde{y}}_1,\ldots, \bm{\tilde{y}}_{\tilde n}  | \bm y]} \mathcal{\bar L}_\lambda\big[\{h(\bm{\tilde y}_i), g(\bm{ \tilde x}_i; \bm{\delta}) \}_{i=1}^{\tilde n}\big].
\end{equation} 
This operation averages the predictive loss over the joint distribution of future or unobserved values $\{h(\bm{\tilde y}_i)\}_{i=1}^n$ at $\mathcal{\widetilde X}$ under model $\mathcal{M}$, and then selects parameters $\bm{\hat \delta}_{\mathcal{A}}$ that minimize this quantity. We define the \emph{parametrized action} $\mathcal{A} \coloneqq (g, \mathcal{P}, \lambda)$
 as a triple consisting of the targeted predictor $g$, the complexity penalty $\mathcal{P}$, and the complexity parameter $\lambda$. 
Since we typically compare among parametrized actions for the same functional $h$, design points $\mathcal{\widetilde X}$, and Bayesian model $\mathcal{M}$, we suppress notational dependence on these terms.



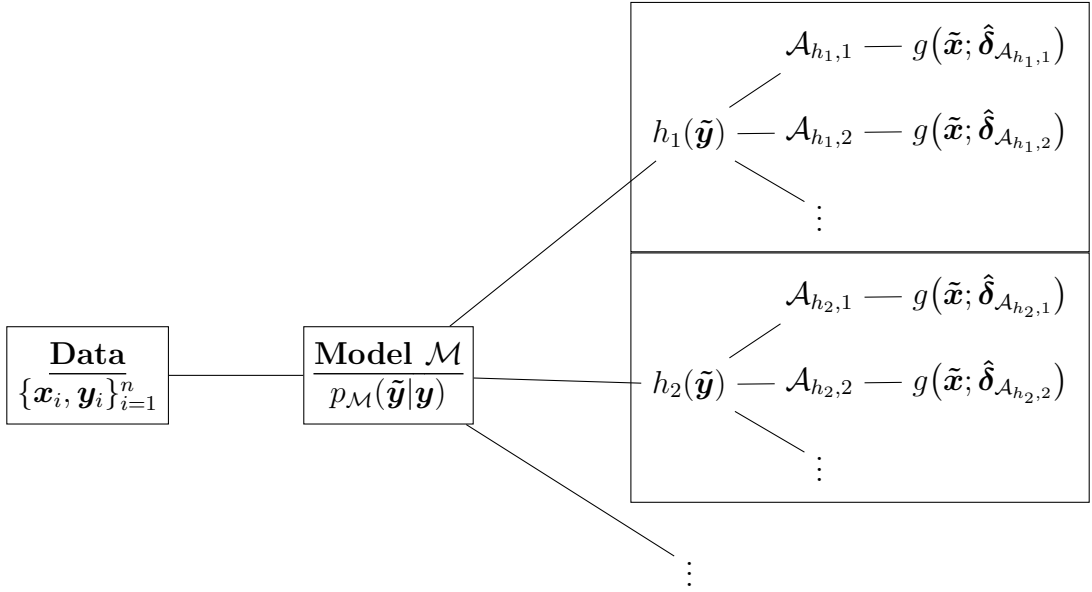
\begin{figure}[ht]
\centering
\begin{forest} 
  [{\bf \ul{Data}} \\ {$\{\bm x_i, \bm y_i\}_{i=1}^n$}, draw, grow = east, align=center, edge={->}
        [{\bf \ul{Model {$\mathcal{M}$}}} \\  {$p_\mathcal{M}(\bm{\tilde y} | \bm y)$} , l = 40mm, draw, grow = east, align=center, s sep = 4mm,
	 [{$\vdots$},  l = 40mm, grow = east, 
	 [,phantom, grow = east,[,phantom, grow = east [,phantom]]]
	 	 [,phantom, grow = east,[,phantom, grow = east [,phantom]]]
	 [,phantom, grow = east,[,phantom, grow = east [,phantom]]]
	 [,phantom, grow = east,[,phantom, grow = east [,phantom]]]
	 ]
	 	 [{$h_2(\bm{\tilde y})$}, l = 40mm, grow = east ,tikz={\node [draw,fit=()(!1)(!ll)] {};}
	 			[{$\vdots$}, l = 10mm] 
				[{$\mathcal{A}_{h_2,2}$}, grow = east, l = 10mm
				[{$g\big(\bm{\tilde x}; \bm{\hat \delta}_{\mathcal{A}_{h_2,2}}\big)$}]
				] 
				[{$\mathcal{A}_{h_2,1}$}, grow = east, l = 10mm
				[{$g\big(\bm{\tilde x}; \bm{\hat \delta}_{\mathcal{A}_{h_2,1}}\big)$}]
				]
	 ] 
	 [{$h_1(\bm{\tilde y})$},  l = 40mm, grow = east, tikz={\node [draw,fit=()(!1)(!ll)] {};}
	 			[{$\vdots$}, l = 10mm] 
				[{$\mathcal{A}_{h_1,2}$}, grow = east, l = 10mm
				[{$g\big(\bm{\tilde x}; \bm{\hat \delta}_{\mathcal{A}_{h_1,2}}\big)$}]
				]
				[{$\mathcal{A}_{h_1,1}$}, grow = east, l = 10mm
				[{$g\big(\bm{\tilde x}; \bm{\hat \delta}_{\mathcal{A}_{h_1,1}}\big)$}]
				]
	 ]
	]
    ]
\end{forest}
\caption{\small 
Given data $\{\bm x_i, \bm y_i\}_{i=1}^n$, a Bayesian model $\mathcal{M}$ is constructed. For each functional $h(\bm{\tilde y})$ and using model $\mathcal{M}$, multiple parametrized actions $\mathcal{A}$ are optimized, evaluated, and compared. The optimal parameters  $\bm{\hat \delta}_{\mathcal{A}}$   are used to compute point predictions $g(\bm{\tilde x}; \bm{\hat \delta}_{\mathcal{A}})$ of $h(\bm{\tilde y})$ at $\bm{\tilde x}$. 
 \label{fig:scheme}}
\end{figure}

The challenge is to produce optimal point prediction parameters $\bm{\hat \delta}_{\mathcal{A}}$ for distinct approximating models $\mathcal{A}$, and subsequently to evaluate and compare the resulting point predictions. A schematic is presented in Figure~\ref{fig:scheme}: given data $\{\bm x_i, \bm y_i\}_{i=1}^n$, a Bayesian model $\mathcal{M}$ is constructed; for each functional $h$, one or more approximations $\mathcal{A}$ are optimized for prediction; point predictions $g(\bm{\tilde x}; \bm{\hat \delta}_{\mathcal{A}})$ are computed for $h(\bm{\tilde y})$ at $\bm{\tilde x}$. The optimal parameters $\bm{\hat \delta}_{\mathcal{A}}$ offer a summary of the posterior (predictive) distribution of model $\mathcal{M}$---akin to posterior expectations, standard deviations, and credible intervals---but specifically targeted to $h$.  

By design, the optimal parameters $\bm{\hat \delta}_{\mathcal{A}}$ depend on the loss function $\mathcal{L}_0$. Generality of $\mathcal{L}_0$ is desirable, but tractability is essential for practical use. 
A natural starting point is squared error loss
$\mathcal{L}_0 \{h(\bm{\tilde y}), g(\bm{ \tilde x}; \bm{\delta}) \} = \big\Vert h(\bm{\tilde y}) - g(\bm{ \tilde x}; \bm{\delta})\big\Vert_2^2$
with generalizations considered below.  In this setting, we identify a representation of the requisite optimization problem \eqref{action} that admits fast and interpretable solutions for a broad class of parametrized actions: 
\begin{theorem}\label{pls-all}
When $\mathbb{E}_{[\bm{\tilde{y}}_i | \bm y]}\big\Vert h(\bm{\tilde y}_i)\big\Vert_2^2 < \infty$ at each $\bm{\tilde x}_i \in \widetilde{\mathcal{X}}, i=1,\ldots,{\tilde n}$, the optimal point prediction parameters in \eqref{action} under squared error loss are
\begin{equation}\label{pls}
\bm{\hat \delta}_{\mathcal{A}} =  \arg\min_{\bm \delta} \left\{{\tilde n}^{-1} \sum_{i=1}^{\tilde n} \big\Vert \bar h_i - g(\bm{\tilde x}_i; \bm{\delta})\big\Vert_2^2+ \lambda \mathcal{P}(\bm \delta)\right\}
\end{equation}
where $\bar h_i \coloneqq \mathbb{E}_{[\bm{\tilde y}_i | \bm y]} h(\bm{\tilde y}_i)$  is the posterior predictive expectation of $h(\bm{\tilde y}_i)$ at $\bm{\tilde x}_i$ under model  $\mathcal{M}$. 
\end{theorem} 
Theorem~\ref{pls-all} establishes an equivalence between the solution to the posterior predictive expected loss \eqref{action} and a penalized least squares criterion, with important computational implications.  
First, estimation of $\bar h_i$ is a standard Bayesian exercise, for example using posterior predictive samples:  $\bar h_i \approx S^{-1} \sum_{s=1}^S h(\bm{\tilde y}_i^{s})$ for $\bm{\tilde y}_i^{s} \sim p_\mathcal{M}(\bm{\tilde y}_i | \bm y)$ at $\bm{\tilde x}_i$. Most commonly,  posterior predictive samples are generated  by iteratively  drawing $\bm \theta^s \sim p_\mathcal{M}(\bm \theta | \bm y)$ from the posterior and  $\bm{\tilde y}_i^s \sim p_\mathcal{M}(\bm{\tilde y}_i | \bm\theta^s)$ from the sampling distribution. Second, the penalized least squares representation in \eqref{pls} implies that the optimal point prediction parameters $\bm{\hat \delta}_\mathcal{A} $ can be computed easily and efficiently for many choices of $\mathcal{A}$ using existing algorithms and software. Third, the optimal parametrized actions produce fast out-of-sample targeted predictions: the prediction of $h(\bm{\tilde y})$ at \emph{any} $\bm{\tilde x}$ is $g(\bm{\tilde x}; \bm{\hat \delta}_\mathcal{A})$, which is quick to compute for many choices of $g$. Lastly, the optimal parameters from \eqref{pls} can be computed simultaneously for many parametrized actions $\mathcal{A}$ and distinct functionals $h$---all based on a single Bayesian model $\mathcal{M}$.

\begin{remark}
Certain choices of $h$, such as binary functionals $h(\bm{\tilde y}) \in \{0,1\}$, are incompatible with squared error loss. In the supplementary material, we discuss generalizations to deviance-based loss functions. 
Importantly, the core attributes of the proposed approach are maintained: computational speed, ease of implementation, and interpretability. 
\end{remark}

We illustrate the utility of this framework with the following examples; an additional example with $h(\bm{\tilde y}) \in \{0,1\}$ is presented in the supplementary material.  

\begin{example}[Linear contrasts]\label{ex-contrasts}
Consider a (multivariate) regression model $\mathbb{E}_{[\bm{y}_i | \bm \theta]} \bm{y}_i = f_{\bm\theta}(\bm x_i)$ for  $\bm{y}_i = (y_{i,1},\ldots,y_{i, m})'$. The linear contrast $h(\bm{\tilde y}) = \bm{C} \bm{\tilde y}$ is often of interest: the matrix $\bm C$ can extract specific components of $\bm{\tilde y}$, evaluate differences between components of $\bm{\tilde y}$, and apply a linear weighting scheme to $\bm{\tilde y}$. For functional data with $y_{i,j} = y_i(\tau_j)$, the linear contrast can target subdomains $ \bm{C} \bm{\tilde y} = \{\tilde y(\tau)\}_{\tau \in \mathcal{S}}$ for $\mathcal{S} \subset \mathcal{T}$ and evaluate derivatives of $\tilde y(\tau)$. In this setting, the predictive target simplifies to the posterior expectation $\bar h =\mathbb{E}_{[\bm \theta | \bm y]}  \{\bm C f_{\bm\theta}(\bm{\tilde x}) \}= \bm C\mathbb{E}_{[\bm \theta | \bm y]} f_{\bm\theta}(\bm{\tilde x})$. Given an estimate $\hat f_{\bm\theta}(\bm{\tilde x}) $ of the posterior expectation of the regression function at $\bm{\tilde x}$, the response variable $\bar h_i \approx \bm C\hat f_{\bm\theta}(\bm{\tilde x}_i)$ needed for \eqref{pls} is easily computable for many choices of $\bm C$. Notably, the predictive expected contrast $ \bm C\hat f_{\bm\theta}(\bm{x}_i)$ is distinct from the empirical contrast $h(\bm y_i) = \bm C \bm y_i$: the former can incorporate shrinkage, smoothness, and other regularization of the regression function $f_{\bm \theta}$ under $\mathcal{M}$. From a single Bayesian model $\mathcal{M}$, multiple parametrized actions $\mathcal{A}$ can be optimized for each contrast $\bm C$.
\end{example}
\begin{example}[Functional data summaries] \label{ex-summaries}
Suppose $h$ is a scalar summary of a curve $\{y(\tau)\}_{\tau \in \mathcal{T}}$,  such as the maximum $h(\bm{\tilde y}) = \max_\tau \tilde y(\tau)$ or the 
point at which the maximum occurs $h(\bm{\tilde y})= \arg\max_\tau \tilde y(\tau)$, 
and let  $\mathcal{M}$ be a Bayesian functional data model (Section~\ref{pa} provides a detailed example). To select variables for optimal linear prediction of $h(\bm{\tilde y})$, we apply Theorem~{\ref{pls-all}} with $g(\bm{\tilde x}; \bm \delta) =  \bm{\tilde x}'\bm\delta$ and an $\ell_1$-penalty, $\mathcal{P}(\bm\delta) = \Vert\bm \delta\Vert_1 = \sum_{j=1}^p |\delta_j|$:
\begin{equation}\label{pls-lasso}
\bm{\hat \delta}_{\mathcal{A}} =  \arg\min_{\bm \delta} \Big\{{\tilde n}^{-1} \sum_{i=1}^{\tilde n} \big\Vert \bar h_i - \bm{\tilde x}_i'\bm\delta \big\Vert_2^2+ \lambda  \Vert\bm \delta\Vert_1\Big\},
\end{equation}
for example using the observed covariates $\mathcal{\widetilde X} = \{\bm x_i\}_{i=1}^n$. 
The optimal  parameters $\bm{\hat \delta}_\mathcal{A}$ are readily computable using existing software, such as \texttt{glmnet} in \texttt{R} \citep{glmnet}.

In practice, we apply an adaptive variant of the $\ell_1$-penalty. Motivated by the adaptive lasso \citep{zou2006adaptive}, \cite{KowalPRIME2020} introduce the penalty $\mathcal{P}(\bm\delta, \bm \theta) =\sum_{j=1}^p \omega_j |\delta_j|$, where $\omega_j = |\beta_j|^{-1}$ and $\beta_j$ are the regression coefficients in a Gaussian linear model  $\mathcal{M}$.  For nonlinear or non-Gaussian models $\mathcal{M}$ and targeted predictions, we use the generalized weights  $\bm \omega = | \bm{\tilde \delta}_0|^{-1}$, where  $\bm{\tilde \delta}_0$ is the $\ell_2$-projection of the predictive variables $h(\bm{\tilde y}_i)$ onto the predictor $g$. 
Bayesian decision analysis requires integration over the unknown $\bm \theta$, so the requisite penalty in \eqref{pls}  becomes the posterior expectation $\overline{\mathcal{P}(\bm \delta)} \coloneqq \mathbb{E}_{[\bm{\tilde y} | \bm y]}\mathcal{P}(\bm \delta, \bm \theta) = \sum_{j=1}^p \hat \omega_j |\delta_j|$ for $\bm{\hat \omega} = \mathbb{E}_{[\bm{\tilde y} | \bm y]} (| \bm{\tilde \delta}_0|^{-1})$, which is estimable using posterior predictive samples.

\end{example}

The parameterized and targeted decision analysis from \eqref{action} features connections with classical decision analysis. Targeted prediction arises in classical decision analysis through the Bayes estimator $\bar h_i = \mathbb{E}_{[\bm{\tilde y}_i | \bm y]} h(\bm{\tilde y}_i)$, which is obtained from Theorem~\ref{pls-all} as a special case:
\begin{corollary}\label{bayes-estimator}
Let $\mathcal{A}_{B} = (g(\bm{\tilde x}; \bm \delta) = \delta(\bm{\tilde x}), \lambda = 0)$ denote an \emph{unrestricted} and \emph{unpenalized} action. The optimal point predictor parameters are $\hat \delta(\bm{\tilde x}_i) = \bar h_i$. 
\end{corollary}
However, action parametrization and penalization are valuable tools: they lend interpretability to the targeted prediction, highlight the balance between accuracy and simplicity, and often produce faster---and more accurate---out-of-sample predictions via $g(\bm{\tilde x}; \bm{\hat \delta}_\mathcal{A})$. 

In some cases, the optimal actions $\bm{\hat \delta}_\mathcal{A}$ can be linked to the underlying model parameters $\bm \theta$, such as when the parameterization $\mathcal{A}$ matches the form of $\mathcal{M}$ and both are linear: 
\begin{corollary}\label{linear-action}
Let $\mathcal{A}_{L} = (g(\bm{\tilde x}; \bm \delta) = \bm{\tilde x}'\bm \delta, \lambda = 0)$ denote a \emph{linear} and \emph{unpenalized} action. For a model $\mathcal{M}$ with $\mathbb{E}_{[\bm{\tilde y}_i | \bm\theta]}h(\bm{\tilde y}_i) = \bm{\tilde x}_i'\bm \theta$ and using the observed design points ${\mathcal{\widetilde X}} = \{\bm{x}_i\}_{i=1}^{n}$, the optimal point predictor parameters are  $\bm{\hat \delta}_{\mathcal{A}_L} =  \mathbb{E}_{\bm \theta | \bm y} \bm \theta$.
\end{corollary}
Corollary~\ref{linear-action} is most familiar when $\mathcal{M}$ is a linear model and $h$ is the identity. By further  allowing $\lambda > 0$ with a sparsity penalty $\mathcal{P}$, we recover the \emph{decoupling shrinkage and selection} approach for Bayesian linear variable selection \citep{hahn2015decoupling}. Similar links to  \cite{woody2019model} can be established for nonlinear regression. 

Despite the potential connections to $\bm \theta$ in certain cases, the parametrized actions are not bound by the parametrization of model $\mathcal{M}$. The full benefits of Theorem~\ref{pls-all} are realized by the simultaneous generality of  the model $\mathcal{M}$, the functionals $h$, and the parametrized actions $\mathcal{A}$. Of course, we can shift the emphasis from prediction toward posterior summarization by replacing the predictive functional $h(\bm{\tilde y})$ with a posterior functional $h(\bm \theta)$, such as $h(\bm \theta) = h\big(\mathbb{E}_{[\bm{\tilde y} | \bm \theta]} \bm{\tilde y}\big)$. However, we prefer the predictive functionals: they correspond to concrete observables that are comparable across Bayesian models \citep{Geisser1993}.

\section{Predictive inference for model determination}\label{out-of-sample}
Decision analysis extracts an optimal $\bm{\hat \delta}_{\mathcal{A}}$ by minimizing a posterior (predictive) expected loss function. However, this optimality is obtained only for a \emph{given} parametrized action  $\mathcal{A}$. 
The key implication of Theorem~\ref{pls-all} is that optimal point predictions can be computed easily and efficiently for \emph{many}  $\mathcal{A}$ (see Figure~\ref{fig:scheme}). To fully exploit these benefits, additional tools are needed to evaluate, compare, and select among the parametrized actions.  

We proceed to evaluate predictive performance {out-of-sample}, which best encapsulates the task of predicting new data. The Bayesian model $\mathcal{M}$ provides {predictive uncertainty quantification} for all evaluations and comparisons. These out-of-sample predictive comparisons serve to identify not only the {best} targeted predictor, but also those targeted predictors that achieve an {acceptable} level of accuracy for out-of-sample prediction. The collection of acceptable targeted predictors illuminates the shared characteristics of near-optimal models, such as the important covariates, the forms of $g$ and $\mathcal{P}$, and the level of complexity needed for  accurate prediction of $h(\bm{\tilde y})$. This approach only requires a Bayesian model $\mathcal{M}$, an \emph{evaluative} loss function $L$, and the design points at which to evaluate the predictions under some $g$.

\subsection{Predictive model evaluation}\label{sec-eval}
The path toward model comparisons and selection begins with evaluation of a single targeted predictor. We proceed nominally using $g(\bm{\tilde x}; \bm{\hat \delta}_{\mathcal{A}})$, 
but note that any point predictor of $h(\bm {\tilde y})$ at $\bm{\tilde x}$ can be used. 
Let $L(\bm z, \bm{\hat z})$ denote the  loss  associated with a prediction $\bm{\hat z}$ when $\bm{ z}$ has occurred. 
We consider both \emph{empirical} and \emph{predictive} versions of the loss: the former uses empirical functionals $\bm z = h(\bm y)$ and relies exclusively on the observed data,  while the latter uses predictive functionals $\bm z = h(\bm{\tilde y})$ and inherits a predictive distribution under $\mathcal{M}$.




Out-of-sample evaluation necessitates a division of the data into \emph{training} and \emph{validation} sets: model-fitting and optimization are restricted to the training data, while predictive evaluations are conducted on the validation data. Dependence on any particular data split is reduced by   repeating this procedure for $K$ randomly-selected splits akin to $K$-fold cross-validation;  we use $K=10$. Let $\mathcal{I}_k \subset \{1,\ldots,n\}$ denote the $k$th validation set, where each data point appears in (at least) one validation set, $\cup_{k=1}^K \mathcal{I}_k = \{1,\ldots,n\}$. We prefer validation sets that are  equally-sized, mutually exclusive, and selected randomly from $\{1,\ldots,n\}$, although other designs are compatible.  Importantly, we do \emph{not} require re-fitting of the Bayesian model $\mathcal{M}$ on each training set, and instead use computationally efficient approximation techniques based on a single fit of $\mathcal{M}$ to the full data (see Section~\ref{approx-oos}).

For each data split $k$, the out-of-sample \emph{empirical} and \emph{predictive} losses are
\begin{equation}
\label{out-loss-k}
\mathbb{L}_{\mathcal{A}}^{out}(k) \coloneqq  \frac{1}{|\mathcal{I}_k|} \sum_{i \in \mathcal{I}_k} L\big\{h(\bm y_i), g(\bm x_i; \bm{\hat \delta}_\mathcal{A}^{-\mathcal{I}_k})\big\}, \quad 
\widetilde{\mathbb{L}}_{\mathcal{A}}^{out}(k)  \coloneqq  \frac{1}{|\mathcal{I}_k|} \sum_{i \in \mathcal{I}_k} L\big\{h(\bm {\tilde y}_i^{-\mathcal{I}_k}), g(\bm x_i; \bm{\hat \delta}_\mathcal{A}^{-\mathcal{I}_k})\big\} 
\end{equation}
respectively, where $\bm{\hat \delta}_\mathcal{A}^{-\mathcal{I}_k}  \coloneqq  \arg\min_{\bm \delta}  \mathbb{E}_{[\bm{\tilde{y}} | \bm y^{-\mathcal{I}_k}]}  \mathcal{\bar L}_\lambda\big[\{h(\bm{\tilde y}_i), g(\bm{ \tilde x}_i; \bm{\delta}) \}_{i \not \in \mathcal{I}_k}\big]$ is optimized only using the training data  $\bm y^{-\mathcal{I}_k} \coloneqq \{\bm y_i\}_{i \not \in \mathcal{I}_k}$,
and similarly $\bm {\tilde y}_i^{-\mathcal{I}_k} \sim p_{\mathcal{M}}(\bm{\tilde y}_i | \bm y^{-\mathcal{I}_k})$ is the predictive variate at $\bm x_i$  conditional only on the training data. Although in-sample versions are available, there is an important distinction between the \emph{out-of-sample} predictive distribution, $p_{\mathcal{M}}(\bm{\tilde y}_i | \bm y^{-\mathcal{I}_k})$, and the \emph{in-sample} predictive distribution, $p_{\mathcal{M}}(\bm{\tilde y}_i | \bm y)$. The in-sample version 
conditions on both the training data $\bm y^{-\mathcal{I}_k}$ and the validation data $\bm y^{\mathcal{I}_k} \coloneqq \{\bm y_i\}_{i \in \mathcal{I}_k}$, which overstates the accuracy and understates the uncertainty for a validation point $i \in \mathcal{I}_k$. The out-of-sample version avoids these issues and more closely resembles most practical prediction problems.

Evaluation of $\mathcal{A}$ is based on the averages of \eqref{out-loss-k} across all data splits, 
$$\mathbb{L}_{\mathcal{A}}^{out} \coloneqq K^{-1} \sum_{k=1}^K \mathbb{L}_{\mathcal{A}}^{out}(k), \quad \widetilde{\mathbb{L}}_\mathcal{A}^{out}  \coloneqq K^{-1} \sum_{k=1}^K\widetilde{\mathbb{L}}_\mathcal{A}^{out}(k).$$
The $K$-fold aggregation averages over two sources of variability in \eqref{out-loss-k}: variability in the training sets $\{\bm x_i, \bm y_i\}_{i \not\in \mathcal{I}_k}$, each of which results in a distinct estimate of the coefficients $\bm{\hat \delta}_\mathcal{A}^{-\mathcal{I}_k}$, and variability in the validation sets $\{\bm x_i, \bm y_i\}_{i \in \mathcal{I}_k}$, which evaluates predictions only at the validation design points $\{\bm x_i\}_{i \in \mathcal{I}_k}$. The contrast between $\mathbb{L}_{\mathcal{A}}^{out}$ and $\widetilde{\mathbb{L}}_{\mathcal{A}}^{out}$ is important:  $\mathbb{L}_{\mathcal{A}}^{out}$ is a point estimate of the 
risk under predictions from $\mathcal{A}$, while $\widetilde{\mathbb{L}}_{\mathcal{A}}^{out}$ provides the distribution of out-of-sample loss under  different \emph{realizations} of the predictive variables $h(\bm{\tilde y}_i)$. Specifically, each $h(\bm y_i)$ for $i \in \mathcal{I}_k$ represents one possible realization of the out-of-sample target variable at $\bm x_i$; the predictive variable $h(\bm {\tilde y}_i^{-\mathcal{I}_k})$ for $\bm {\tilde y}_i^{-\mathcal{I}_k} \sim p_{\mathcal{M}}(\bm{\tilde y}_i | \bm y^{-\mathcal{I}_k})$ expresses the distribution of possible realizations according to $\mathcal{M}$. The predictive loss $\widetilde{\mathbb{L}}_{\mathcal{A}}^{out}$ incorporates this distributional information for out-of-sample predictive uncertainty quantification.

\subsection{Predictive model selection}\label{sec-select}
The out-of-sample empirical and predictive losses, $\mathbb{L}_{\mathcal{A}}^{out}$ and $\widetilde{\mathbb{L}}_\mathcal{A}^{out}$, respectively, provide the ingredients needed to  compare and select among targeted predictors. Predictive quantities have proven useful for Bayesian model selection; see \cite{vehtari2012survey} for a thorough review. Our goal is not only to identify the {most accurate} predictor, but also to gather those targeted predictors that achieve an {acceptable} level of accuracy. In doing so, we introduce a Bayesian representation of the \emph{Rashomon effect}, which observes that for many practical applications, many approaches can achieve adequate predictive accuracy \citep{breiman2001statistical}. 



The proposed notion of ``acceptable" accuracy is defined relative to the most accurate targeted predictor, $
\mathcal{A}_{min} \coloneqq \arg\min_{\mathcal{A} \in \mathbb{A}}  \mathbb{L}_\mathcal{A}^{out}
$, which minimizes out-of-sample empirical loss as in classical $K$-fold cross-validation. The set $\mathbb{A}$ may include different forms for $g$ and $\mathcal{P}$ and usually will include a path of $\lambda$ values for each $(g, \mathcal{P})$ pair. We prefer relative rather than absolute accuracy because it directly references an empirically attainable accuracy level. 

For any two actions $\mathcal{A}, \mathcal{A}' \in \mathbb{A}$,   let 
$\widetilde{\mathbb{D}}_{\mathcal{A}, \mathcal{A}'}^{out} \coloneqq 100\times(\widetilde{\mathbb{L}}_\mathcal{A}^{out} - \widetilde{\mathbb{L}}_{\mathcal{A}'}^{out})/\widetilde{\mathbb{L}}_{\mathcal{A}'}^{out}$
 be the percent increase in out-of-sample predictive loss from $\mathcal{A}$ to $\mathcal{A}'$. We seek all parametrized actions $\mathcal{A}$ that perform within a margin $\eta \ge 0\%$ of the best model, $\widetilde{\mathbb{D}}_{\mathcal{A}, \mathcal{A}_{min}}^{out} < \eta\%$, with  probability at least $\varepsilon \in [0,1]$. The margin $\eta$ acknowledges that near-optimal performance---especially for simple models---is often sufficient, while the probability level $\varepsilon$ incorporates predictive uncertainty. In concert,  $\eta$ and  $\varepsilon$ provide domain-specific and model-informed leniency for admission into a set of acceptable predictors. We formally define the set of \emph{acceptable predictors} as follows:
\begin{definition}
The \emph{set of acceptable predictors} is  $\Lambda_{\eta, \varepsilon} \coloneqq \big\{\mathcal{A}\in \mathbb{A}: \mathbb{P}_\mathcal{M}\big(\widetilde{\mathbb{D}}_{\mathcal{A}, \mathcal{A}_{min}}^{out} < \eta \big) \ge \varepsilon \big\}$, where $\widetilde{\mathbb{D}}_{\mathcal{A}, \mathcal{A}_{min}}^{out} \coloneqq 100\times(\widetilde{\mathbb{L}}_\mathcal{A}^{out} - \widetilde{\mathbb{L}}_{\mathcal{A}_{min}}^{out})/\widetilde{\mathbb{L}}_{\mathcal{A}_{min}}^{out}$. 
\end{definition}
The probability $\mathbb{P}_\mathcal{M}$ is estimated using out-of-sample predictive draws under model $\mathcal{M}$ (see Section~\ref{approx-oos}). Each set $\Lambda_{\eta, \varepsilon}$  is nonempty, since  $\mathcal{A}_{min} \in \Lambda_{\eta, \varepsilon}$ for all $\eta, \varepsilon$, and nested: $\Lambda_{\eta', \varepsilon'} \subseteq \Lambda_{\eta, \varepsilon}$ for any $\eta' \ge \eta$ or $\varepsilon' \le \varepsilon$, so increasing $\eta$ or decreasing $\varepsilon$ can expand the set of acceptable predictors. The special case of sparse Bayesian linear regression was considered in \cite{KowalPRIME2020}.  With similar intentions, \cite{Tulabandhula2013} and \cite{Semenova2019} define a \emph{Rashomon set} of predictors for which the in-sample empirical loss is within a margin $\eta$ of the best predictor. By comparison, $\Lambda_{\eta, \varepsilon}$  uses out-of-sample criteria for evaluation and incorporates predictive uncertainty via the Bayesian model $\mathcal{M}$.

The set of acceptable predictors also can be constructed using prediction intervals: 
\begin{lemma}\label{accept-lemma}
A predictor $\mathcal{A}$ is acceptable, $\mathcal{A} \in  \Lambda_{\eta, \varepsilon}$, if and only if  there exists a lower $(1 - \varepsilon)$  posterior prediction interval for  $\widetilde{\mathbb{D}}_{\mathcal{A}, \mathcal{A}_{min}}^{out} $ that  includes $\eta$.
\end{lemma}
Viewed another way, $\mathcal{A}$ is \emph{not} acceptable if the lower $1 - \varepsilon$ predictive interval for $\widetilde{\mathbb{D}}_{\mathcal{A}, \mathcal{A}_{min}}^{out} $ excludes $\eta$. From this perspective, \emph{unacceptable} predictors are those $\mathcal{A}$ for which there is insufficient predictive probability (under $\mathcal{M}$) that the out-of-sample  accuracy of $\mathcal{A}$ is within a certain margin of the best predictor. This definition is similar to the confidence sets of \cite{Lei2019}, which exclude any $\mathcal{A}$ for which the null hypothesis that $\mathcal{A}$ produces best predictive risk is rejected.  \cite{Lei2019} relies on a customized bootstrap procedure, which adds substantial computational burden to the model-fitting and cross-validation procedures. By comparison, acceptable predictor sets are derived entirely from the predictive distribution of $\mathcal{M}$ and accompanied by fast and accurate approximation algorithms (see Section~\ref{approx-oos}).

Among acceptable predictors, we highlight the simplest one.  For fixed $(g, \mathcal{P})$, the simplest predictor has the largest complexity penalty:  $\lambda_{\eta, \varepsilon} \coloneqq \max\{\lambda: (g, \mathcal{P}, \lambda) \in \Lambda_{\eta, \varepsilon}\}.$
When $\mathcal{P}$ is a sparsity penalty such as \eqref{pls-lasso}, the simplest acceptable predictor contains the smallest set of covariates needed to (nearly) match the predictive accuracy of the best predictor---which may itself be $\mathcal{A}_{min}$. 
Selection based on $\lambda_{\eta, \varepsilon} $ resembles the \emph{one-standard-error rule} (e.g., \citealp{Hastie2009}), which selects the simplest predictor for which the out-of-sample empirical loss is within one standard error of the best predictor. Instead, $\lambda_{\eta, \varepsilon}$ uses the out-of-sample predictive loss with posterior uncertainty quantification inherited  from $\mathcal{M}$.

\subsection{Fast approximations for out-of-sample predictive evaluation}\label{approx-oos}
The primary hurdle for out-of-sample predictive evaluations is computational: they require computing  $\bm{\hat \delta}_\mathcal{A}^{-\mathcal{I}_k}$ and sampling $\bm {\tilde y}_i^{-\mathcal{I}_k} \sim p_{\mathcal{M}}(\bm{\tilde y}_i | \bm y^{-\mathcal{I}_k})$ for each data split $k=1,\ldots,K$. 
Re-fitting $\mathcal{M}$ on each training set $\{\bm x_i, \bm y_i\}_{i \not\in \mathcal{I}_k}$ is impractical and in many cases computationally infeasible. 
To address these challenges, we develop efficient approximations that require only a \emph{single fit} of the Bayesian model $\mathcal{M}$ to the data---which is already necessary for standard posterior inference. Specifically, we use a sampling-importance resampling (SIR) algorithm with the full posterior predictive distribution as a proposal for the relevant out-of-sample predictive distributions. The subsequent results focus on squared error loss, but adaptations to other loss functions are straightforward. 

To obtain $\bm{\hat \delta}_\mathcal{A}^{-\mathcal{I}_k}$, we equivalently represent the optimal action as in Theorem~\ref{pls-all}:
\begin{equation}\label{loo-el}
\bm{\hat \delta}_\mathcal{A}^{-\mathcal{I}_k} =  \arg\min_{\bm \delta}
\Big\{ (n - |\mathcal{I}_k|)^{-1} \sum_{j \not\in \mathcal{I}_k} \big\Vert \bar h_j^{-\mathcal{I}_k} - g({\bm x}_{j}; {\bm \delta})\big\Vert_2^2 + \lambda \mathcal{P}(\bm \delta)\Big\}
\end{equation}
where $\bar h_j^{-\mathcal{I}_k} = \mathbb{E}_{[\bm{\tilde y}_j | \bm y^{-\mathcal{I}_k}]} h(\bm{\tilde y}_j)$  is the out-of-sample point prediction at $\bm x_j$. As such, \eqref{loo-el} is easily solvable for many choices of $\mathcal{A}$: all that is required is an estimate of  $\bar h_j^{-\mathcal{I}_k}$ for each $j \not \in \mathcal{I}_k$ in the training set. We estimate this quantity using importance sampling. Proposals $\{\bm{\tilde y}_j^s\}_{s=1}^S \sim p_\mathcal{M}( \bm{\tilde y}_j | \bm y)$ are generated from the full predictive distribution by 
sampling  $\{\bm \theta^s\}_{s=1}^S \sim p_\mathcal{M}(\bm \theta | \bm y)$ from the full posterior and  $\{\bm{\tilde y}_j^s\}_{s=1}^S \sim p_\mathcal{M}( \bm{\tilde y}_j | \bm \theta^s)$ from the likelihood. The full data posterior $p_\mathcal{M}(\bm \theta | \bm y)$ serves as a proposal for the training data posterior $p_\mathcal{M}(\bm \theta | \bm y^{-\mathcal{I}_k})$ with importance weights $w_k^s \propto 1/p(\bm y^{\mathcal{I}_k} | \bm \theta^s)$, with further factorization under  conditional independence. The target can be estimated using $\bar h_j^{-\mathcal{I}_k} \approx \sum_{s=1}^S w_k^s  h(\bm{\tilde y}_j^s)$ or based on SIR sampling. In some cases, it is beneficial to regularize the importance weights  \citep{Ionides2008,vehtari2015pareto},  but our empirical results remain unchanged with or without regularization. 
Successful variants of this approach exist for Bayesian model selection \citep{gelfand1992model} and evaluating prediction distributions \citep{vehtari2012survey}.



SIR provides a mechanism for sampling $\bm {\tilde y}_i^{-\mathcal{I}_k} \sim p_{\mathcal{M}}(\bm{\tilde y}_i | \bm y^{-\mathcal{I}_k})$ using the importance weights $\{w_k^s\}_{s=1}^S$, which in turn provides out-of-sample predictive draws of  $\widetilde{\mathbb{L}}_{\mathcal{A}}^{out}$ and  $\widetilde{\mathbb{D}}_{\mathcal{A}, \mathcal{A}'}^{out}$ for any actions $\mathcal{A}, \mathcal{A}' \in \mathbb{A}$. The idea is to obtain the proposal samples $\{\bm {\tilde y}_j^s\}_{s=1}^S \sim p_{\mathcal{M}}(\bm{\tilde y}_j | \bm y)$ from the full posterior distribution and then subsample from $\{\bm {\tilde y}_j^s\}_{s=1}^S$ without replacement based on the corresponding importance weights $\{w_k^s\}_{s=1}^S$. The full SIR algorithm details are provided in the  supplementary material.  

\section{Simulation study}\label{sims}
We evaluate the selection capabilities and predictive accuracy of the proposed techniques using synthetic data. For targeted prediction, these evaluations must be directed toward a particular \emph{functional} of the response variable. Specifically, we generate functional data $\{Y_i^*(\tau): \tau \in [0,1]\}$ such that the argmax of each function, $\tau_i^* \coloneqq \arg\max_\tau Y_i^*(\tau)= h(Y_i^*)$, 
 is linearly associated with a subset of covariates, $\tau_i^* = \bm x_i'\bm \beta^*$. The covariates are correlated and mixed continuous and discrete: we draw $x_{i,j}$ from marginal standard normal distributions with  $\mbox{Cor}(x_{i,j}, x_{i,j'}) = (0.75)^{|j - j'|}$ and binarize half of these $p$ variables, $x_{i,j} \leftarrow \mathbb{I}\{x_{i,j} \ge 0\}$. The continuous covariates are centered and scaled to sample standard deviation 0.5. For the true  coefficients $\{\beta_j^*\}_{j=1}^p$, we randomly select 5\% for $\beta_j^* = 1$, 5\% for $\beta_j^* = -1$, and leave the remaining values at zero with the exception of the intercept, $\beta_0^*=1$. The coefficients $\{\beta_j^*\}_{j=0}^p$ are rescaled such that $\tau_i^* = \bm x_i'\bm \beta^* \in [0.2, 0.8]$ to ensure that the argmax occurs away from the boundary; see the supplementary material. The true functions are computed as $Y_i^*(\tau) = a_{0,i} + a_{1,i}\tau - (a_{1,i} + a_{2,i}) (\tau - \tau_i^*)_+$, where  $a_{0,i} \stackrel{iid}{\sim}N(0,1)$, $a_{1,i}, a_{2,i} \stackrel{iid}{\sim} \chi_5^2$, and $(x)_+ \coloneqq x \mathbb{I}\{x \ge 0\}$. By construction, $Y_i^*$ is piecewise linear and concave with a single breakpoint, $\tau_i^* = \arg\max_\tau Y_i^*(\tau) $, and therefore $h(Y_i^*)  = \bm x_i' \bm\beta^*$. 
Finally, the observed  data $\bm y_i$ are generated by adding Gaussian noise to $Y_i^*(\tau)$ at $m$ equally-spaced points with a root signal-to-noise ratio of 5. Example figures are provided in the supplementary material.



The synthetic data-generating process is repeated 100 times for $p=50$ covariates, $m=200$ observation points, and varying sample sizes $n \in\{75, 100, 500\}$. 
For each simulated dataset $\{\bm x_i, \bm y_i\}_{i=1}^n$, we compute the   posterior and predictive distributions under the Bayesian function-on-scalars regression model of \cite{kowal2019bayesianfosr}, which models a linear association between the functional data response and the scalar covariates. We emphasize that this model $\mathcal{M}$ does \emph{not} reflect the true data-generating process, yet our targeted predictions are derived from the predictive distribution under $\mathcal{M}$. We consider linear actions $g(\bm{\tilde x}; \bm \delta) = \bm{\tilde x}'\bm\delta$ with the adaptive $\ell_1$-penalty from  Example~\ref{ex-summaries} and computed using \texttt{glmnet} in \texttt{R} \citep{glmnet}. In this case, the set of parametrized actions $\mathbb{A}$ is determined by the path of $\lambda$ values, which control the sparsity of the linear action $\bm \delta$.  For benchmark comparisons, we use the adaptive lasso \citep{zou2006adaptive} and projection predictive feature selection \citep{piironen2018projective} on the empirical functionals $\{\bm x_i, h(\bm y_i)\}_{i=1}^n$. Model sizes were selected using 10-fold cross-validation.  Implementation of \cite{piironen2018projective} uses the \texttt{projpred} package in \texttt{R}; for the requisite Bayesian linear model, we assume double exponential priors for the linear coefficients, but results are unchanged for Gaussian and t-priors. 



To validate the proposed definition of acceptable predictor sets, we investigate a simple yet important question: does the true model belong to  $\Lambda_{\eta, \varepsilon}$? Specifically, we determine whether the true set of active variables $\{j: \beta_j^* \ne 0\}$ matches the set of active variables for \emph{any} acceptable predictor $\mathcal{A}\in\Lambda_{\eta, \varepsilon}$. This task is challenging:  we do not assume knowledge of the active variables, so the true model only belongs to $\Lambda_{\eta, \varepsilon}$ when it is both correctly identified  along the $\lambda$ path and correctly evaluated by $\widetilde{\mathbb{D}}_{\mathcal{A}, \mathcal{A}'}^{out}$. Correct identification is only satisfied when \emph{all} and \emph{only} the true active variables $\{j: \beta_j^* \ne 0\}$ are nonzero according to $\mathcal{A}$.

\begin{figure}[h!]
\begin{center}
\includegraphics[width=.49\textwidth]{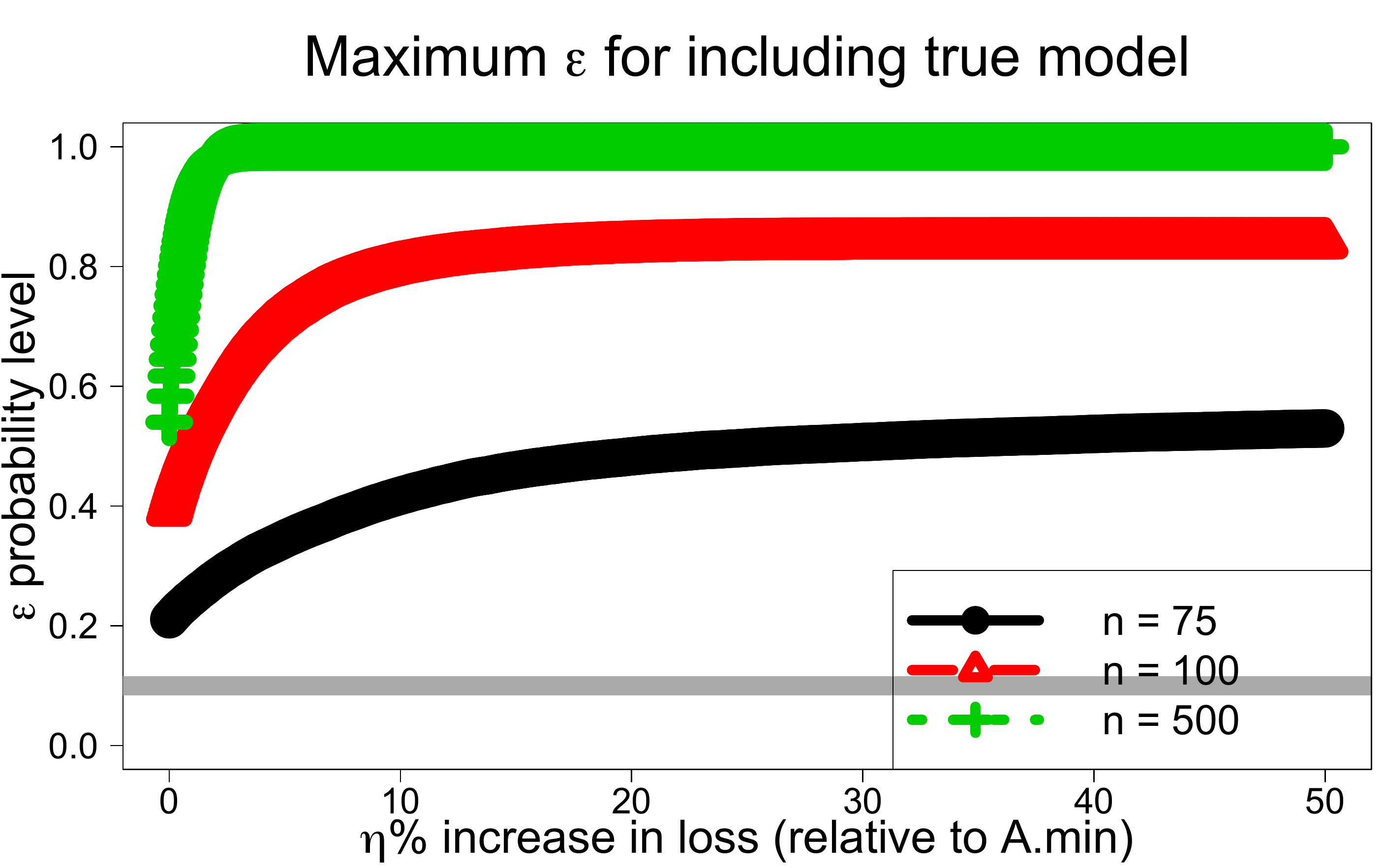}
\caption{\small The maximum probability level $\varepsilon_{max}(\eta)$ for which the true model is acceptable, $\mathcal{A}^* \in \Lambda_{\eta, \varepsilon}$, across values of $\eta$. For any smaller probability level $\varepsilon' \le  \varepsilon_{max}(\mathcal{A}^*)$, the true model remains acceptable: $\mathcal{A}^* \in \Lambda_{\eta, \varepsilon'}$. The horizontal gray line is $\varepsilon = 0.1$.
\label{fig:select}}
\end{center}
\end{figure}

For this task, we compute $\varepsilon_{max}(\eta) \coloneqq \mathbb{P}_\mathcal{M}\big(\widetilde{\mathbb{D}}_{\mathcal{A}^*, \mathcal{A}_{min}}^{out} < \eta \big)$, which is the maximum probability level for which the true model $\mathcal{A}^*$ is acceptable. The margin $\eta$ corresponds to the percent increase in loss relative to  $\mathcal{A}_{min}$.  By design, $\mathcal{A}^* \in \Lambda_{\eta, \varepsilon'}$ remains acceptable for any smaller probability level $\varepsilon' \le  \varepsilon_{max}(\eta)$. Most important, we set $\varepsilon_{max}(\eta) = 0$ if $\mathcal{A}^*$ is not on the $\lambda$ path. For each simulated dataset, we compute $\varepsilon_{max}(\eta)$ for a grid of $\eta\%$ values. The results averaged across 100 simulations are in Figure~\ref{fig:select}. Naturally, $\varepsilon_{max}(\eta)$ uniformly increases with the sample size for each value of $\eta$. When $\eta = 0$, the average maximum probability levels are $\varepsilon_{max}(0) \in \{0.21, 0.39, 0.54\}$ for $n \in\{75, 100, 500\}$, respectively, which suggests that a cutoff of $\varepsilon = 0.1$ is capable of capturing the true model even when zero margin is allowed. Notably, $\varepsilon_{max}(\eta)$ does \emph{not} converge to one as $\eta$ increases for the smaller sample sizes $n \in \{75, 100\}$. The reason is simple: if $\mathcal{A}^*$ is not discovered along the $\lambda$ path, then $\varepsilon_{max}(\eta) = 0$ by definition---regardless of the choice of $\eta$. This result demonstrates the importance of the set of predictors \emph{under consideration} $\mathbb{A}$, which here is determined entirely by the selected variables in the \texttt{glmnet} solution path.

\begin{figure}[h!]
\begin{center}
\includegraphics[width=.49\textwidth]{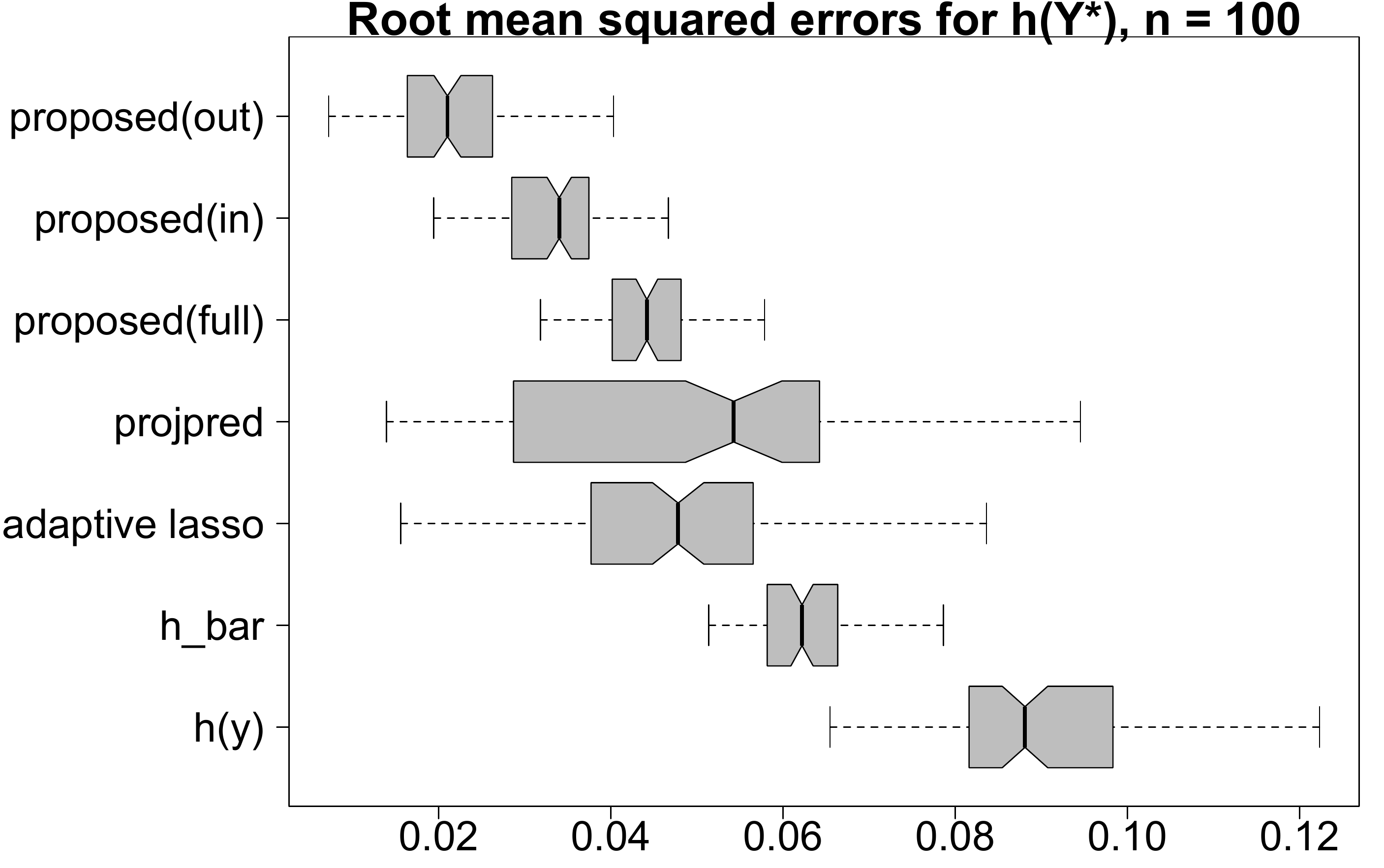}
\includegraphics[width=.49\textwidth]{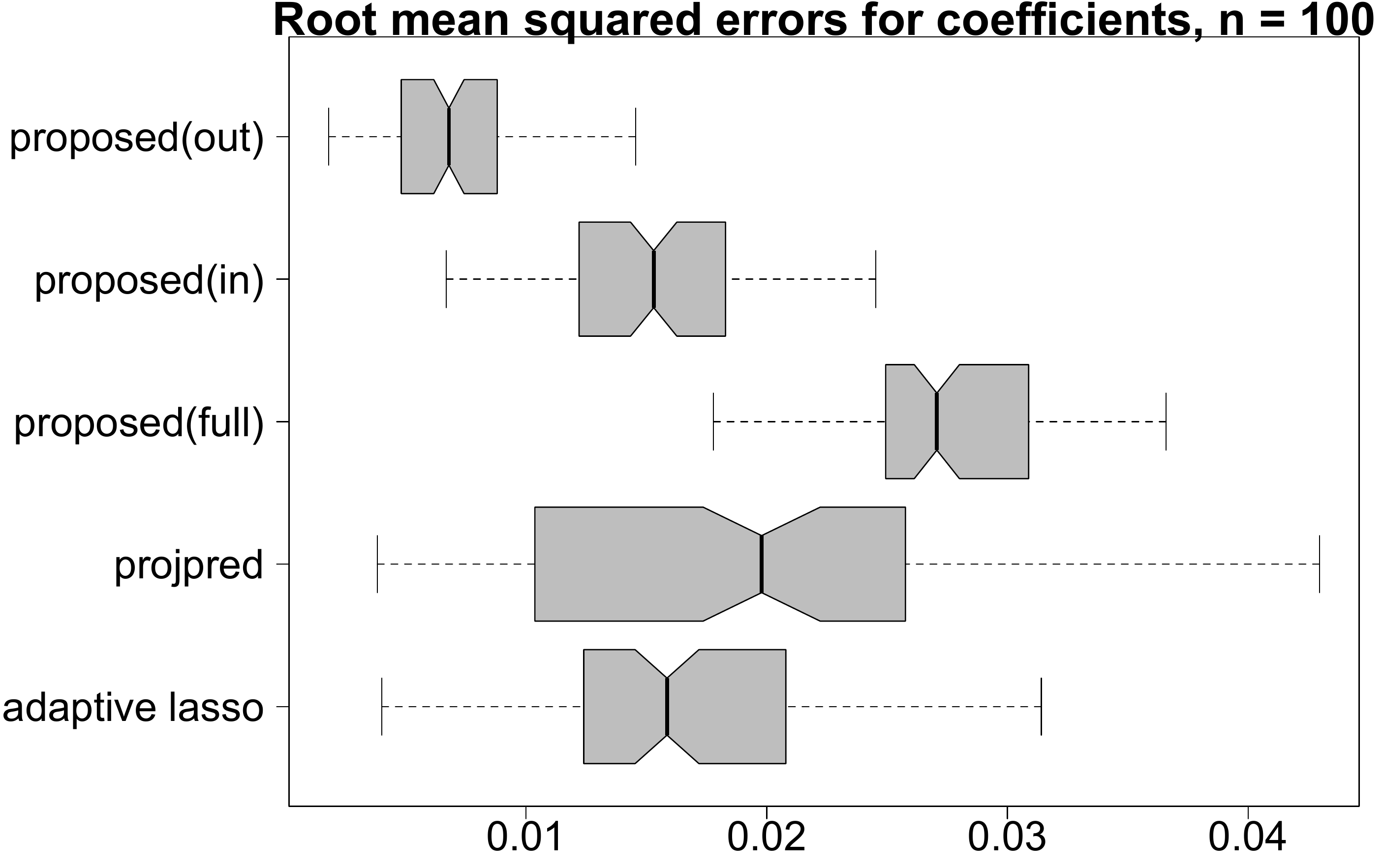}

\includegraphics[width=.49\textwidth]{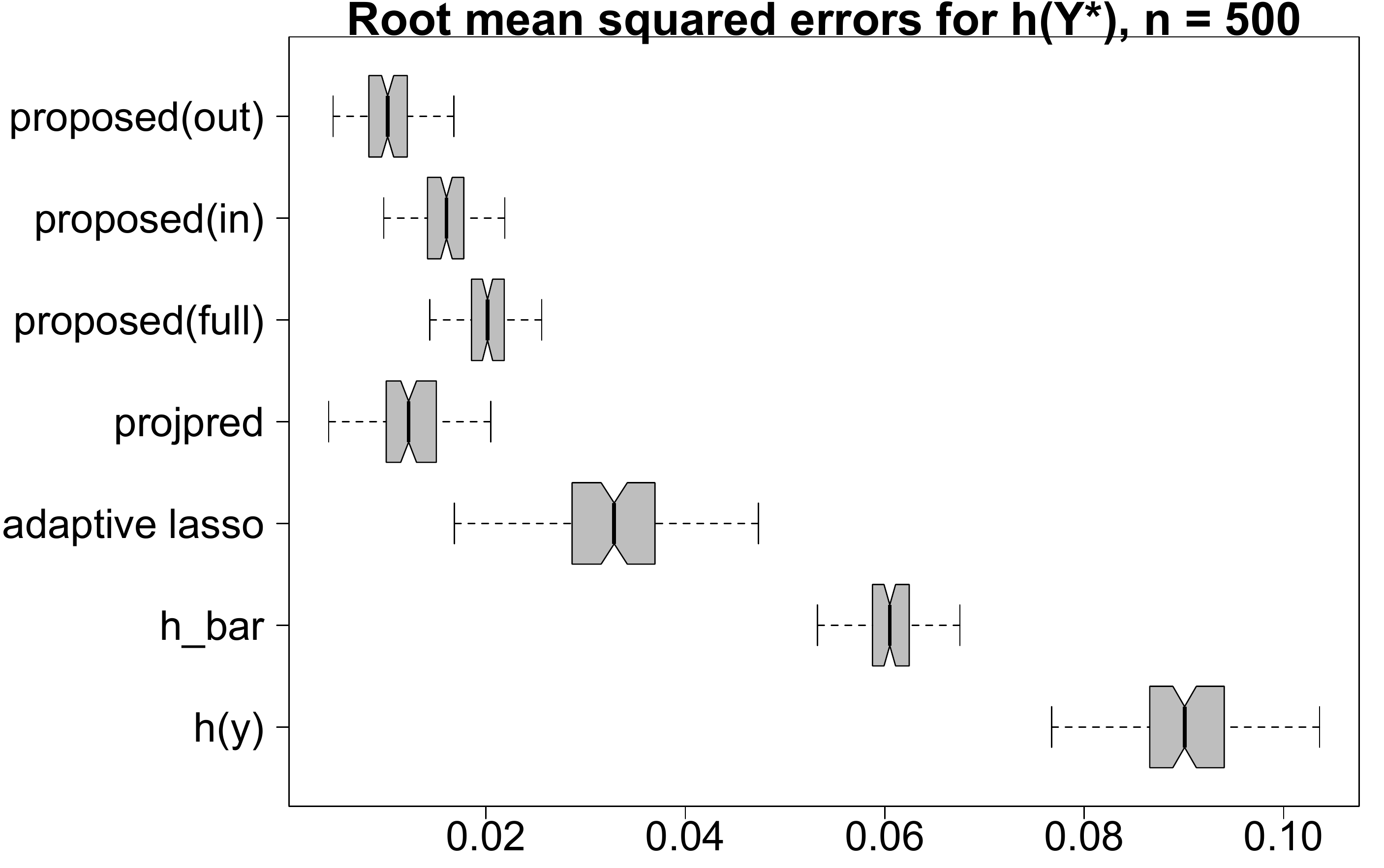}
\includegraphics[width=.49\textwidth]{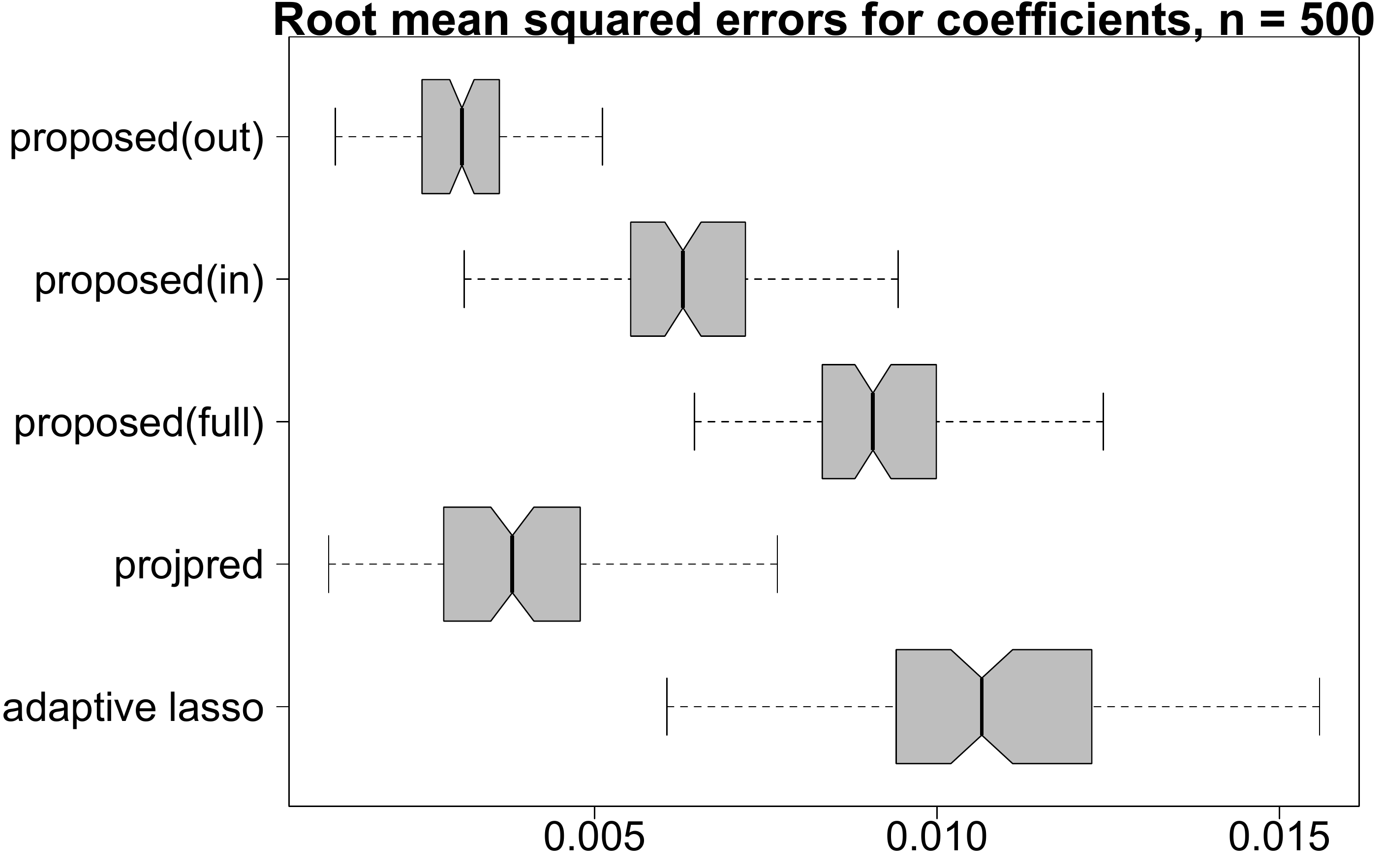}
\caption{\small RMSEs for the true functionals $h(Y^*)$ ({\bf left}) and the true regression coefficients $\bm \beta^*$  ({\bf right}) for $n=100$ ({\bf top}) and $n=500$ ({\bf bottom}) across 100 simulated datasets. Non-overlapping notches indicate significant differences between medians. The parametrized actions with out-of-sample selection are most accurate for prediction and estimation. 
\label{fig:rmse}}
\end{center}
\end{figure}

Next, we evaluate point predictions of $h(Y_i^*)$ and estimates of $\bm \beta^*$ using root mean squared errors (RMSEs). The parametrized actions $\bm{\hat \delta}_\lambda$ and point predictions  $g(\bm{\tilde x}; \bm{\hat \delta}_\lambda) = \bm{\tilde x}'\bm{\hat \delta}_\lambda$ are computed for multiple choices of $\lambda$:  the simplest acceptable predictor $\lambda = \lambda_{\eta, \varepsilon}$ with $\eta =0$ and $\varepsilon = 0.1$ (\texttt{proposed(out)});  the analogous choice of $\lambda$ based on \emph{in-sample} evaluations (\texttt{proposed(in)}); and the unpenalized linear action with $\lambda = 0$ (\texttt{proposed(full)}). For comparisons, we include the aforementioned \texttt{adaptive lasso} and \texttt{projpred}, the point predictions $\bar h_i$ under model $\mathcal{M}$ (\texttt{h\_bar}; see Corollary~\ref{bayes-estimator}),  and the empirical functionals $h(\bm y_i)$ (\texttt{h(y)}).  
The results are in Figure~\ref{fig:rmse}. In summary, clear  improvements in targeted prediction are obtained by  (i) fitting to $h(\bm{\tilde y}_i)$ (via $\bar h_i$) rather than $h(\bm{y}_i)$, (ii) including covariate information, (iii) incorporating penalization or variable selection, and (iv) selecting the complexity $\lambda$ based on out-of-sample criteria. The targeted actions $\mathcal{A}$ vastly outperform the model $\mathcal{M}$ predictions---even though each $\mathcal{A}$ is based entirely on the predictive distribution from $\mathcal{M}$. Lastly, the accurate estimation of the linear coefficients is important: the estimates $\bm{\hat \delta}_\lambda$ describe the partial linear effects of each $x_j$ on targeted prediction of $h(\bm{\tilde y})$.

The supplementary material includes additional comparisons. Marginal variable selection is evaluated based on true positive and false negative rates, with \texttt{proposed(out)} offering the best performance among these methods. Results for high dimensional data with $p > n$ ($n=200, p = 500$ and $n=100, p = 200$) confirm the prediction and estimation advantages of the proposed approach. Sensitivity to $\varepsilon \in \{0.05, 0.10, 0.20, 0.50\}$ is also studied for prediction, estimation, and selection. Lastly, we evaluate the robustness in predictive accuracy among these methods. Specifically, we consider the setting in which the distribution of the covariates differs significantly between the training and validation datasets. The parametrized actions offer superior targeted predictions, especially for small to moderate sample sizes. 

\section{Physical activity data analysis}\label{pa}
We apply targeted prediction to study physical activity (PA) data from the National Health and Nutrition Examination Survey (NHANES). NHANES is a large  survey conducted by the Centers for Disease Control to study the health and wellness of the U.S. population. We analyze data from the 2005-2006 cohort, which features minute-by-minute PA data measured by hip-worn accelerometers (see Figure~\ref{fig:max-pa}). To date, the 2005-2006 cohort is the most recent publicly available NHANES PA data. These data are high-resolution and empirical measurements of PA, and offer an opportunity to study intraday activity profiles. 

PA has been linked to all-cause mortality  not only in \emph{total daily activity} \citep{Schmid2015} but also via other \emph{functionals} that describe activity behaviors \citep{Fishman2016,Smirnova2019}. 
Our goal is to construct targeted predictions that more accurately predict and explain   the defining characteristics of PA. Specifically, we consider the following functionals $h(\bm{\tilde y})$ for intraday PA $\bm{\tilde y} = (\tilde y(\tau_1),\ldots, \tilde y(\tau_m))'$ at times-of-day $\tau_1,\ldots,\tau_m$:
\begin{table}[ht]
\centering
\small
\begin{tabular}{ c | c | c | c | c | c  }
\texttt{avg} &  \texttt{tlac} &\texttt{sd} &  \texttt{sedentary} &  \texttt{max} &  \texttt{argmax} \\ \hline
 $\int \tilde y(\tau) \, d\tau \!$&  $\int \log\{\tilde y(\tau) + 1\} \, d\tau\!$ & $\Vert \tilde y -  \int \tilde y(\tau) \, d\tau \Vert_{L^2}\!$ &    $\int \mathbb{I}\{\tilde y(\tau) \le 100\} \, d\tau\!$ &  $\max_\tau \tilde y(\tau)\!$ &  $ \arg\max_\tau \tilde y(\tau)\!$  
\end{tabular}
\end{table}

\vspace{-2mm}
\noindent where \texttt{avg} captures average daily activity, \texttt{tlac} is the total log activity count and measures moderate activity \citep{Varma2017}, \texttt{sd} targets the intraday variability in PA, \texttt{sedentary} computes the amount of time below a low activity threshold, \texttt{max} is the peak activity level, and \texttt{argmax} is the time of peak activity. In addition, we include a binary indicator of absolute inactivity during sleeping hours:  $\texttt{zeros(1am-5am)} \coloneqq \mathbb{I}\{ \tilde y(\tau) = 0 \mbox{ for all } \tau \in [\mbox{1am}, \mbox{5am}]\}$. Individuals with $\texttt{zeros(1am-5am)} = 1$ likely removed the accelerometer during sleep in accordance with the NHANES instructions. Since we omit subjects with insufficient accelerometer wear time ($<10$ hours), individuals with $\texttt{zeros(1am-5am)} = 1$ are active at other times of the day. 


The PA data are accompanied by 
demographic variables (age, gender, body mass index (BMI), race, and education level), behavioral attributes (smoking status and alcohol consumption), self-reported comorbidity factors (diabetes, coronary heart disease (CHD), congestive heart failure,
cancer, and stroke), and lab measurements (total cholesterol, HDL cholesterol, systolic blood pressure). Data pre-processing generally follows \cite{Leroux2019} using the \texttt{R} package \texttt{rnhanesdata}. We consider individuals aged 50-85 without mobility problems and without missing covariates. 
The continuous covariates are centered and scaled to sample standard deviation 0.5.

In accordance with the schematic in Figure~\ref{fig:scheme}, targeted predictive decision analysis begins with a Bayesian model $\mathcal{M}$. Since the PA data are intraday activity counts, we use a count-valued functional regression model based on the simultaneous transformation and rounding (STAR) framework of \cite{Kowal2020a}. STAR formalizes the popular approach of \emph{transforming} count data prior to applying Gaussian models, but includes a latent \emph{rounding} layer to produce a valid count-valued data-generating process. STAR models can capture zero-inflation, over- and under-dispersion, and boundedness or censoring, and provide a path for adapting continuous data models and algorithms to the count data setting.


For each individual, we aggregate PA across all available days (at least three and at most seven days per subject) in five-minute bins. Let $y_{i,j}$  and $y_{i,j}^{tot}$ and denote the {average} and {total} PA, respectively, for subject $i$ at time $\tau_j$, where $i = 1,\ldots, n = 1012$ and $j = 1,\ldots, m=288$. Total PA is count-valued and will serve as the input for the STAR model, while all subsequent functionals and predictive distributions use average PA. Model $\mathcal{M}$ is the following:
\begin{align}
\label{star-round}
y_{i,j}^{tot} &= \texttt{round}(y_{i,j}^*), \quad z_{i,j}^* = \texttt{transform}(y_{i,j}^*)  \\
\label{star-basis}
z_{i,j}^* &= \bm b'(\tau_j) \bm \theta_i + \sigma_\epsilon \epsilon_i, \quad \epsilon_i \stackrel{iid}{\sim} t_\nu(0,1)\\
\label{star-reg}
\theta_{i, \ell} &= \bm x_i' \bm \alpha_\ell + \sigma_{\gamma_i}\gamma_{i,\ell}, \quad \gamma_{i,\ell} \stackrel{iid}{\sim}N(0,1)
\end{align}
with $ \alpha_{\ell,j} \stackrel{indep}{\sim} N(0, \sigma_{\alpha_j}^2)$ and 
$\sigma_\epsilon^{-2}, \sigma_{\gamma_i}^{-2}, \sigma_{\alpha_j}^{-2} \stackrel{iid}{\sim} \mbox{Gamma}(0.01, 0.01)$. In \eqref{star-round}, \texttt{round} maps the latent continuous data $y_{i,j}^*$ to $\{0,1,\ldots,\infty\}$, while \texttt{transform} maps $y_{i,j}^*$ to $\mathbb{R}$ for continuous data modeling. We use $\texttt{round}(t) = \lfloor t \rfloor$ for $t > 0$ and $\texttt{round}(t) = 0$ for $t \le 0$, so $y_{i,j}^{tot} = 0$ whenever $y_{i,j}^* < 0$, and set $\texttt{transform}(t) = 2 (\sqrt{t} -1)$ in the Box-Cox family. In the functional regression levels \eqref{star-basis}-\eqref{star-reg}, $\bm b$ is a vector of spline basis functions with basis coefficients $\bm \theta_i$ for subject $i$ and $\bm \alpha_\ell$ is the vector of regression coefficients for each basis coefficient $\ell$. The spline basis is reparametrized to orthogonalize  $\bm b$ and diagonalize the prior variance of the basis coefficients, which justifies the assumption of independence across basis coefficients in \eqref{star-reg}. Heavy-tailed innovations ($\nu = 3$) are introduced to model large  spikes in PA.

Posterior inference is conducted based on 5000 samples from a Gibbs sampler after discarding a burn-in of 5000 iterations; the algorithm is detailed in the supplementary material. Posterior predictive diagnostics (see the supplementary material) demonstrate adequacy  of $\mathcal{M}$ for the functionals of interest. These results are insensitive to $\nu$, but alternative choices of \texttt{transform} (e.g., $\log t$) or $\bm b$ (e.g., wavelets) produce inferior results. 


Targeted predictions for each functional were constructed using a linear action  $g(\bm{\tilde x}; \bm \delta) = \bm{\tilde x}'\bm\delta$ with an adaptive $\ell_1$-penalty (see Example~\ref{ex-summaries}). Trees were also considered but were not competitive. The set of parametrized actions $\mathbb{A}$ is given by the path of $\lambda$ values computed using \texttt{glmnet} in \texttt{R} \citep{glmnet}: we highlight the simplest acceptable action $\lambda= \lambda_{0, 0.1}$  (\texttt{proposed(out)}) and the unpenalized linear action $\lambda = 0$  (\texttt{proposed(full)}). For comparison, we fit an adaptive lasso to $\{\bm x_i, h(\bm y_i)\}_{i=1}^n$ for each $h$. Squared error loss is used for all but \texttt{zeros(1am-5am)} which uses cross-entropy. In the supplementary material, we consider quadratic effects for age and BMI and pairwise interactions for each of age and BMI with ethnicity, gender, the behavioral attributes, and the self-reported comorbidity factors.

\begin{figure}[h!]
\begin{center}
\includegraphics[width=.49\textwidth]{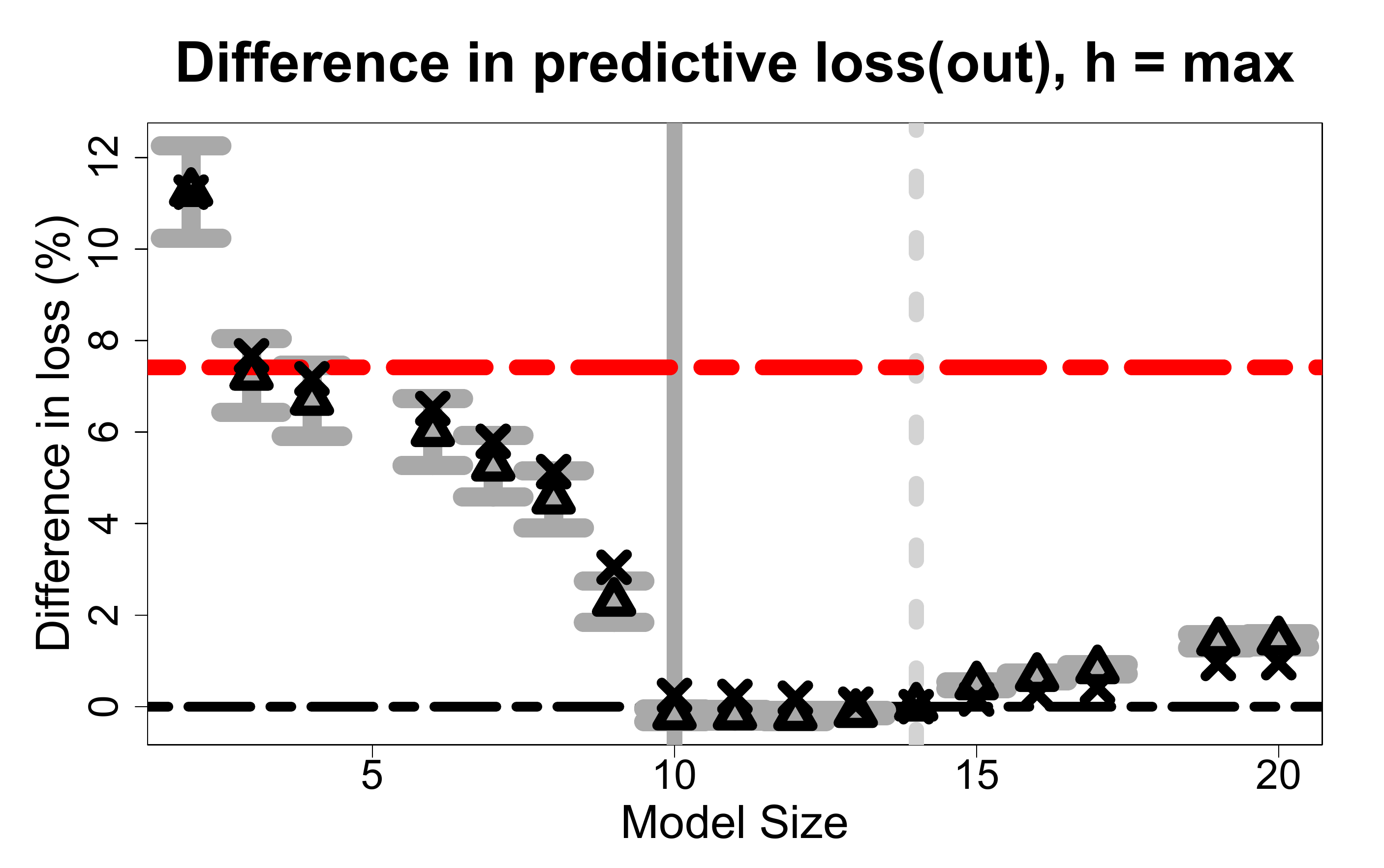}
\includegraphics[width=.49\textwidth]{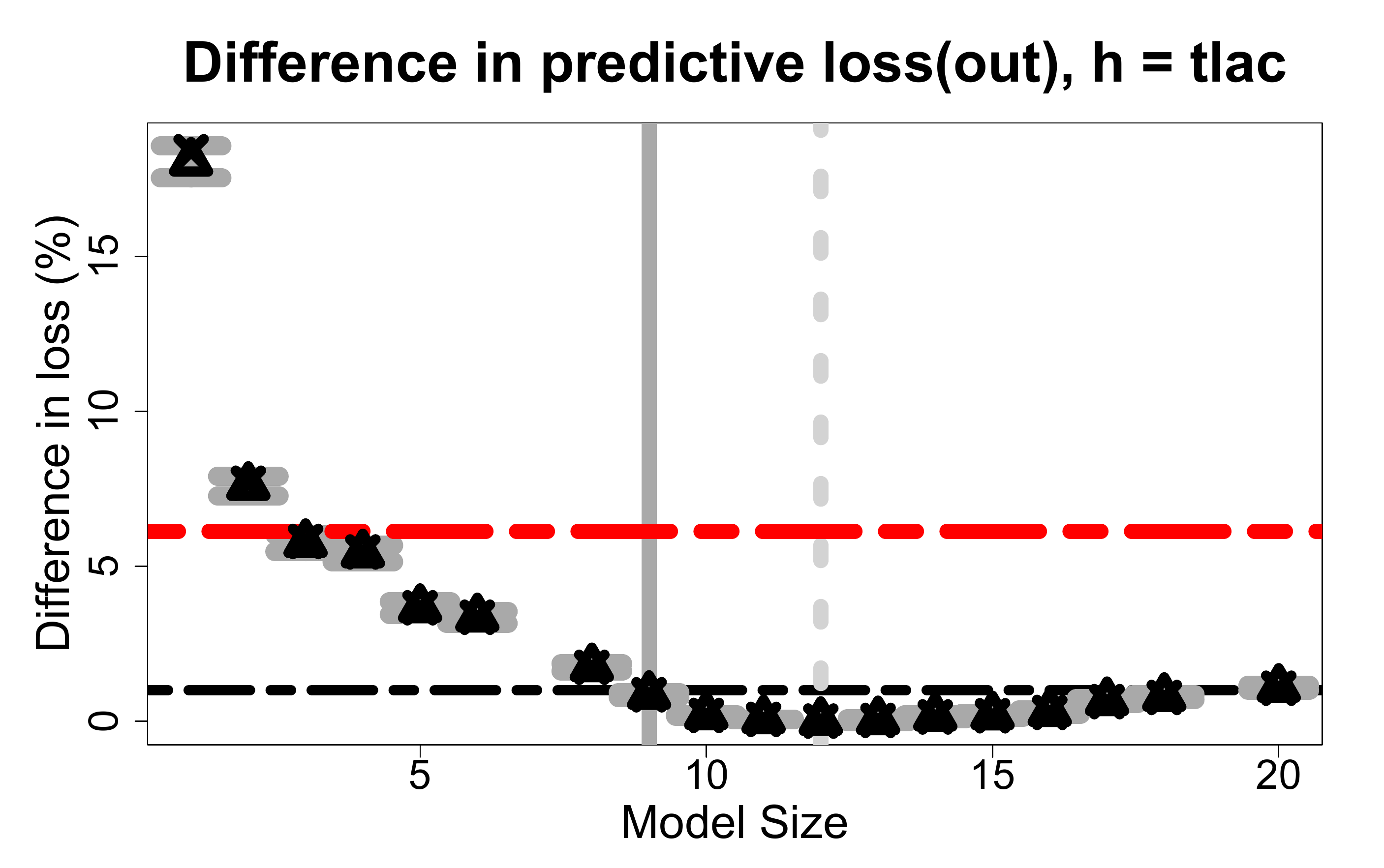}

\includegraphics[width=.49\textwidth]{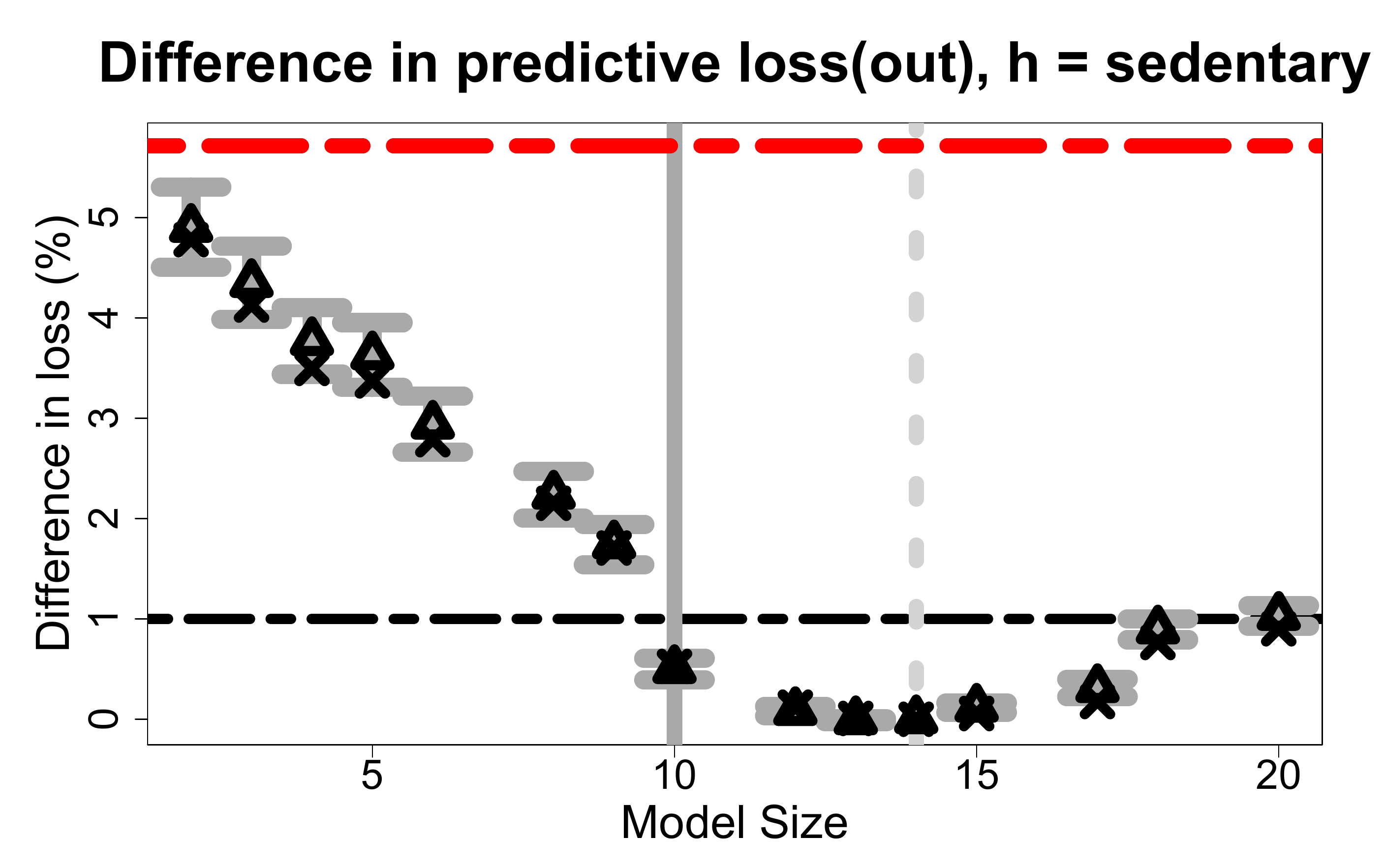}
\includegraphics[width=.49\textwidth]{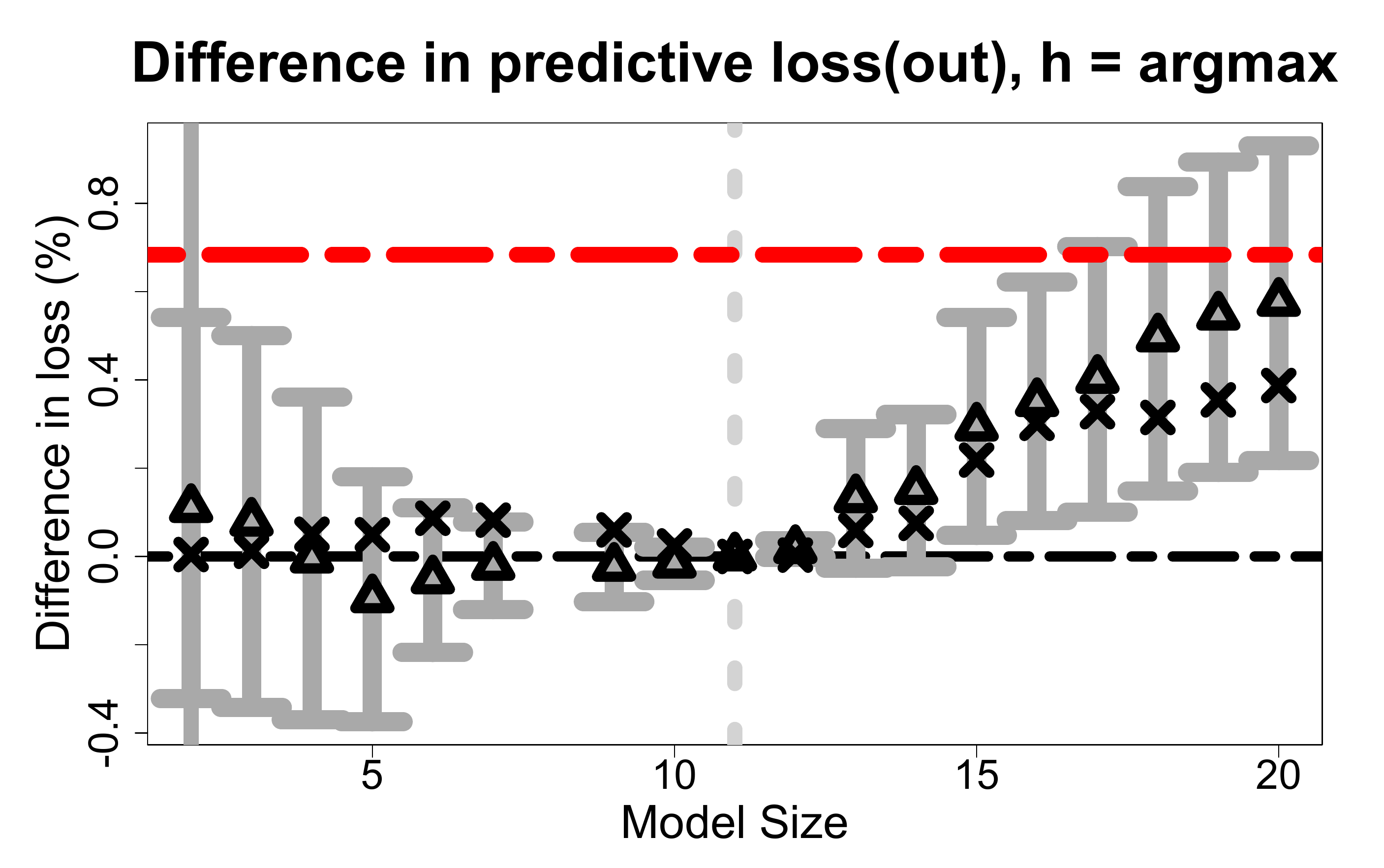}
\caption{\small Approximate out-of-sample squared error loss for sparse linear actions targeted to each functional. Results are presented for each size as a percent increase in loss relative to $\mathcal{A}_{min}$. The predictive expectations  (triangles) and 80\% intervals (gray bars) are included with the empirical relative loss for each model size (x-marks) and the adaptive lasso (red lines). The horizontal black lines denote the choices of $\eta$ and the vertical lines denote $\lambda_{\eta, 0.1}$ (solid)  and $\mathcal{A}_{min}$ (dashed). %
\label{fig:loss-out}}
\end{center}
\end{figure}

The targeted predictions are evaluated out-of-sample using the approximations from Section~\ref{approx-oos}. For each functional $h$ and complexity $\lambda$---which indexes the number of nonzero elements in $\bm{\hat \delta}_{\mathcal{A}}$---Figure~\ref{fig:loss-out} presents the percent increase in predictive and empirical loss relative to the best predictor $\mathcal{A}_{min}$. The measures of vigorous PA (\texttt{avg}, \texttt{sd}, and \texttt{max}) produce nearly identical results, so we only include \texttt{max} here; \texttt{avg} and \texttt{sd} are in the supplement. The predictive expectations align closely with the empirical values, which suggests that model $\mathcal{M}$ is adequate for these predictive metrics.

For each functional, we obtain optimal or near-optimal predictions with only about 10 covariates with better accuracy than the adaptive lasso. Many of the selected covariates are shared among functionals: age, BMI, gender, ethnicity, HDL cholesterol, and CHD are selected for all but \texttt{argmax}, while smoking status (\texttt{avg}, \texttt{sd}, \texttt{max}), diabetes (\texttt{avg}, \texttt{sd}, \texttt{sedentary},  \texttt{max}), and total cholesterol (\texttt{tlac}, \texttt{sedentary}) appear as well. The functionals measuring vigorous PA agree on the selected variables, including negative effects for diabetes and smoking. Most distinct is \texttt{argmax}: while $\mathcal{A}_{min}$ includes 11 covariates, the predictive uncertainty quantification from $\widetilde{\mathbb{D}}_{\mathcal{A}, \mathcal{A}_{min}}^{out}$ indicates that linear predictors with as few as one covariate (ethnicity) are acceptable.  These covariates are simply not linearly predictive of \texttt{argmax}: the difference between $\mathcal{A}_{min}$ and any other $\mathcal{A} \in \mathbb{A}$ is less than 1\%.

Robustness to the choice of $\eta$ is also illustrated in Figure~\ref{fig:loss-out}. We select $\eta = 0\%$ for \texttt{max} and \texttt{argmax} and $\eta = 1\%$ for \texttt{tlac} and \texttt{sedentary}, which highlights the purpose of $\eta$: by allowing $\eta > 0$, we can obtain targeted predictors with fewer covariates. By comparison, increasing the margin to $\eta = 1\%$ for \texttt{max} and \texttt{argmax} does not change the smallest acceptable predictor.









To validate the approximations in Figure~\ref{fig:loss-out}, we augment the analysis with a truly out-of-sample prediction evaluation. For each of 20 training/validation splits, model $\mathcal{M}$ and the adaptive lasso are fit to the training data and sparse linear actions are targeted to each $h$. We emphasize that this exercise is computationally intensive: the MCMC for model $\mathcal{M}$ requires about 30 minutes per 10000 iterations (using R on a MacBook Pro, 2.8 GHz Intel Core i7), so repeating the model-fitting process 20 times is extremely slow. Comparatively, the approximations used for Figure~\ref{fig:loss-out} compute in under two seconds.

\begin{figure}[h!]
\begin{center}
\includegraphics[width=.32\textwidth]{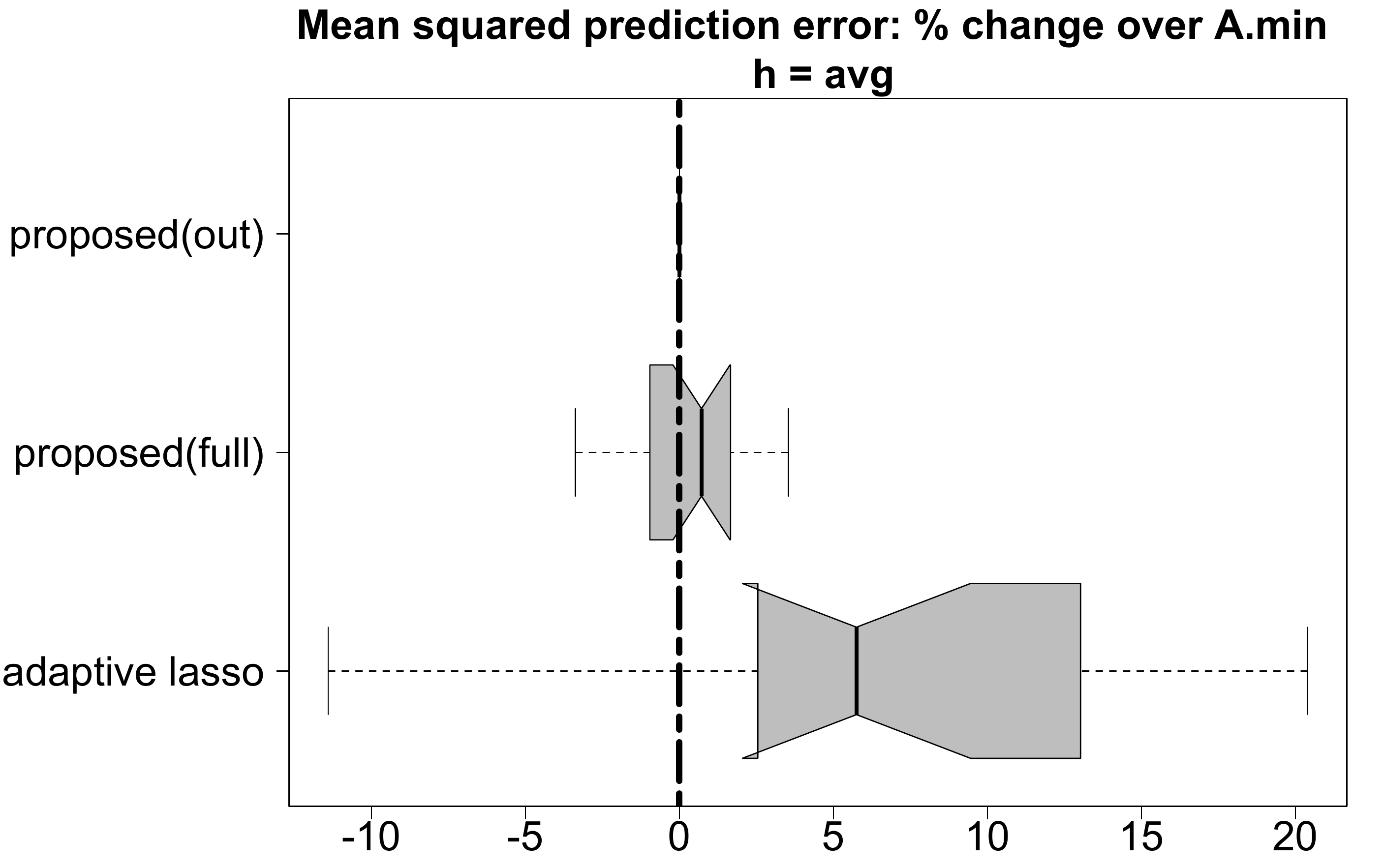}
\includegraphics[width=.32\textwidth]{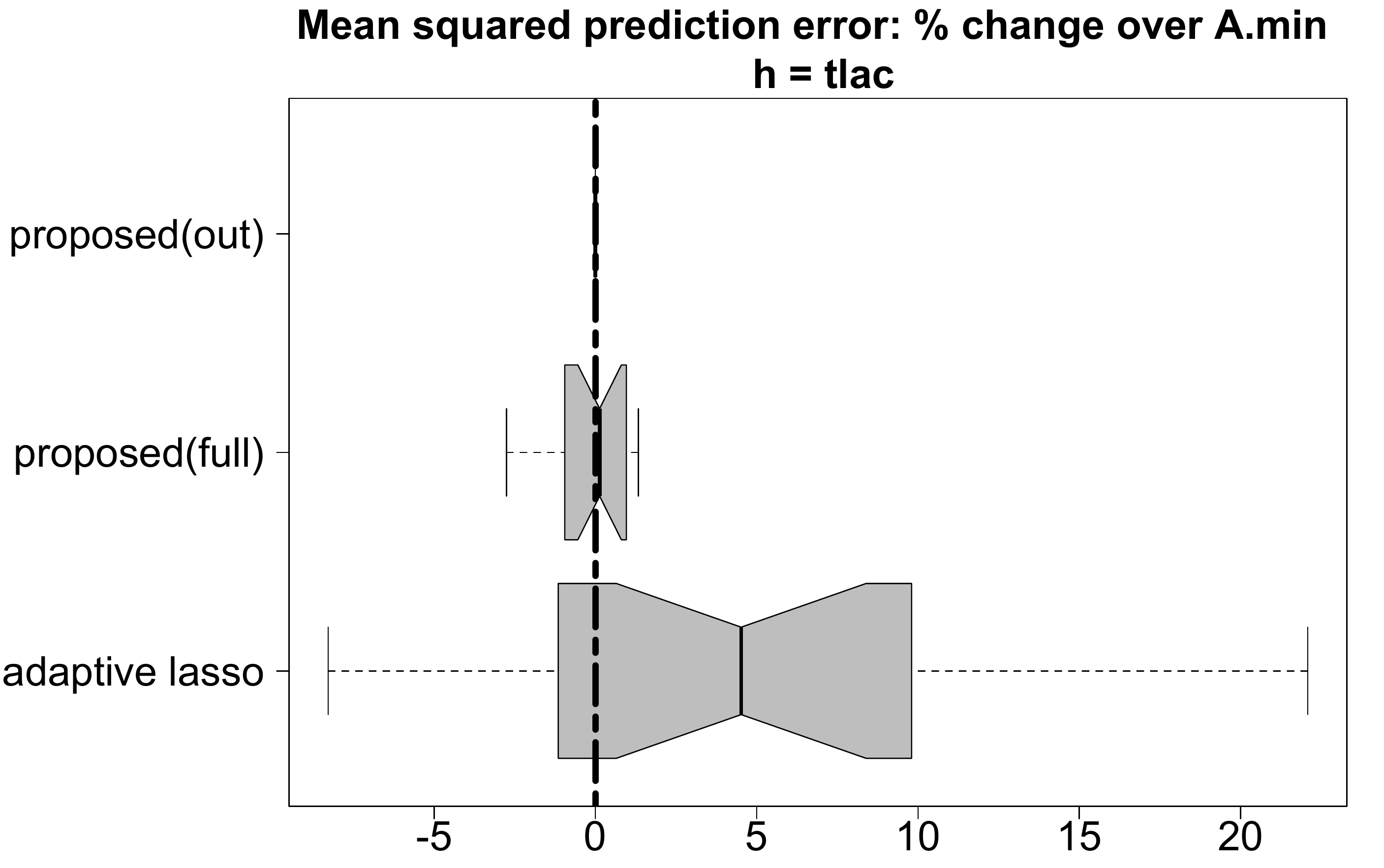}
\includegraphics[width=.32\textwidth]{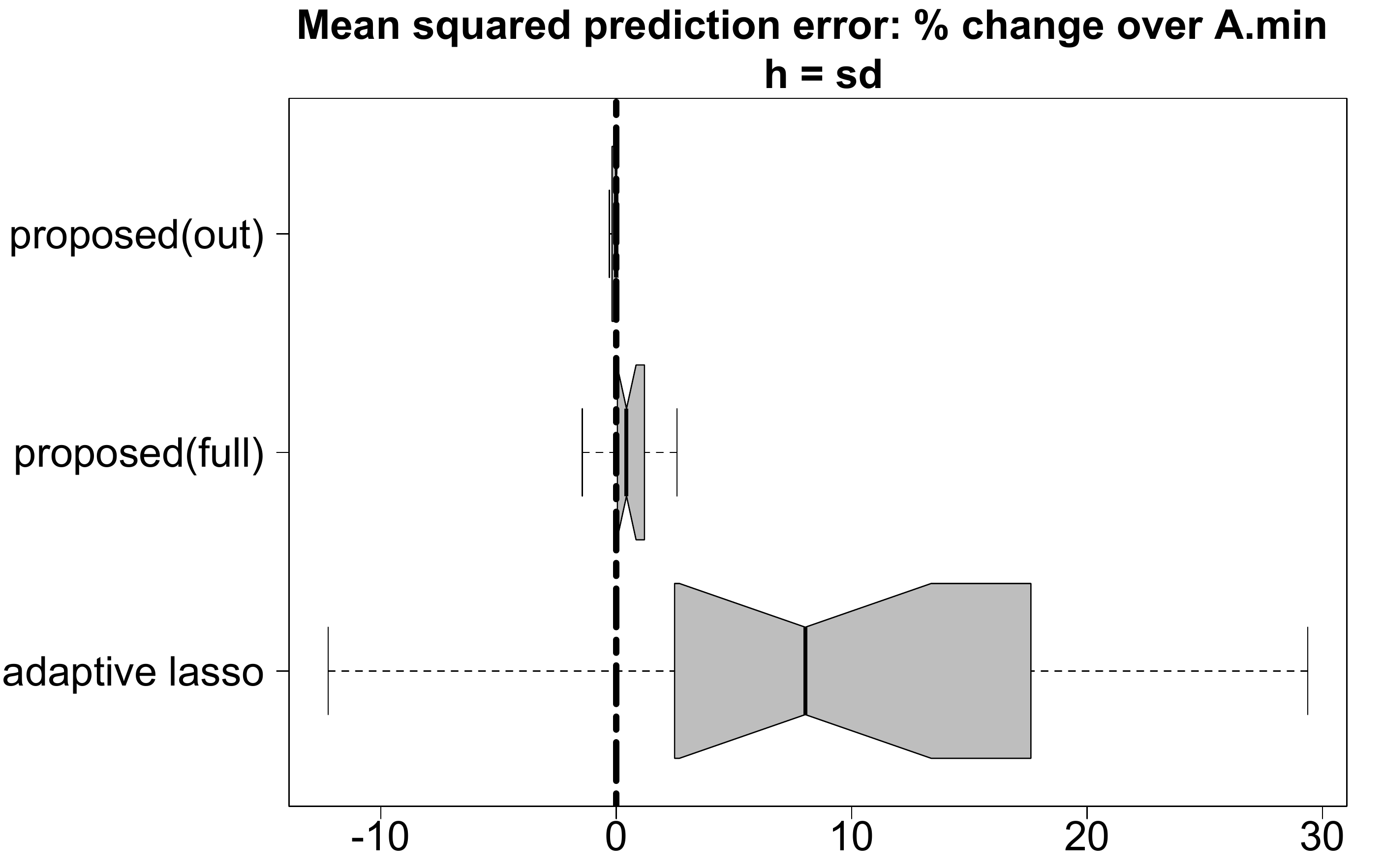}

\includegraphics[width=.32\textwidth]{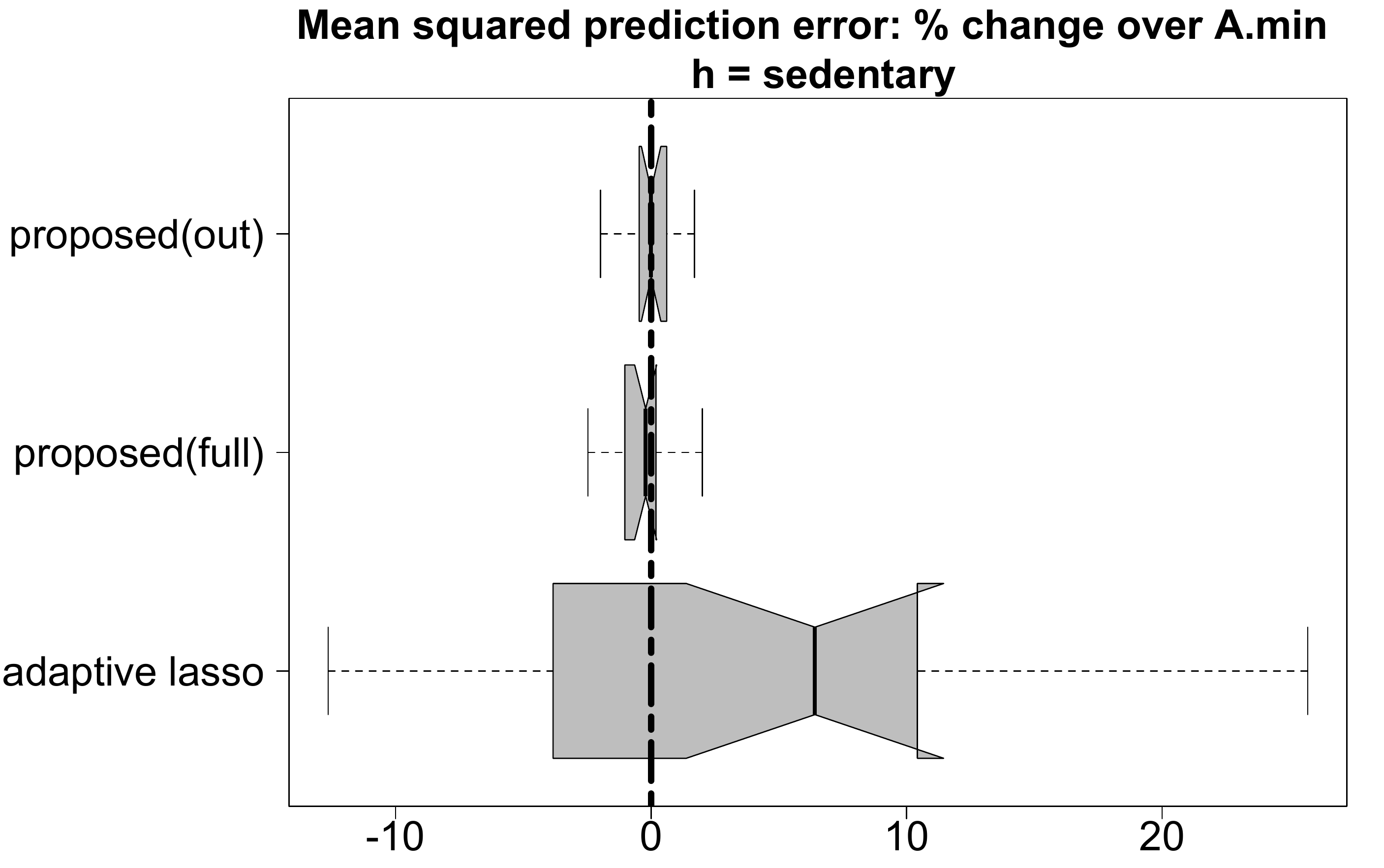}
\includegraphics[width=.32\textwidth]{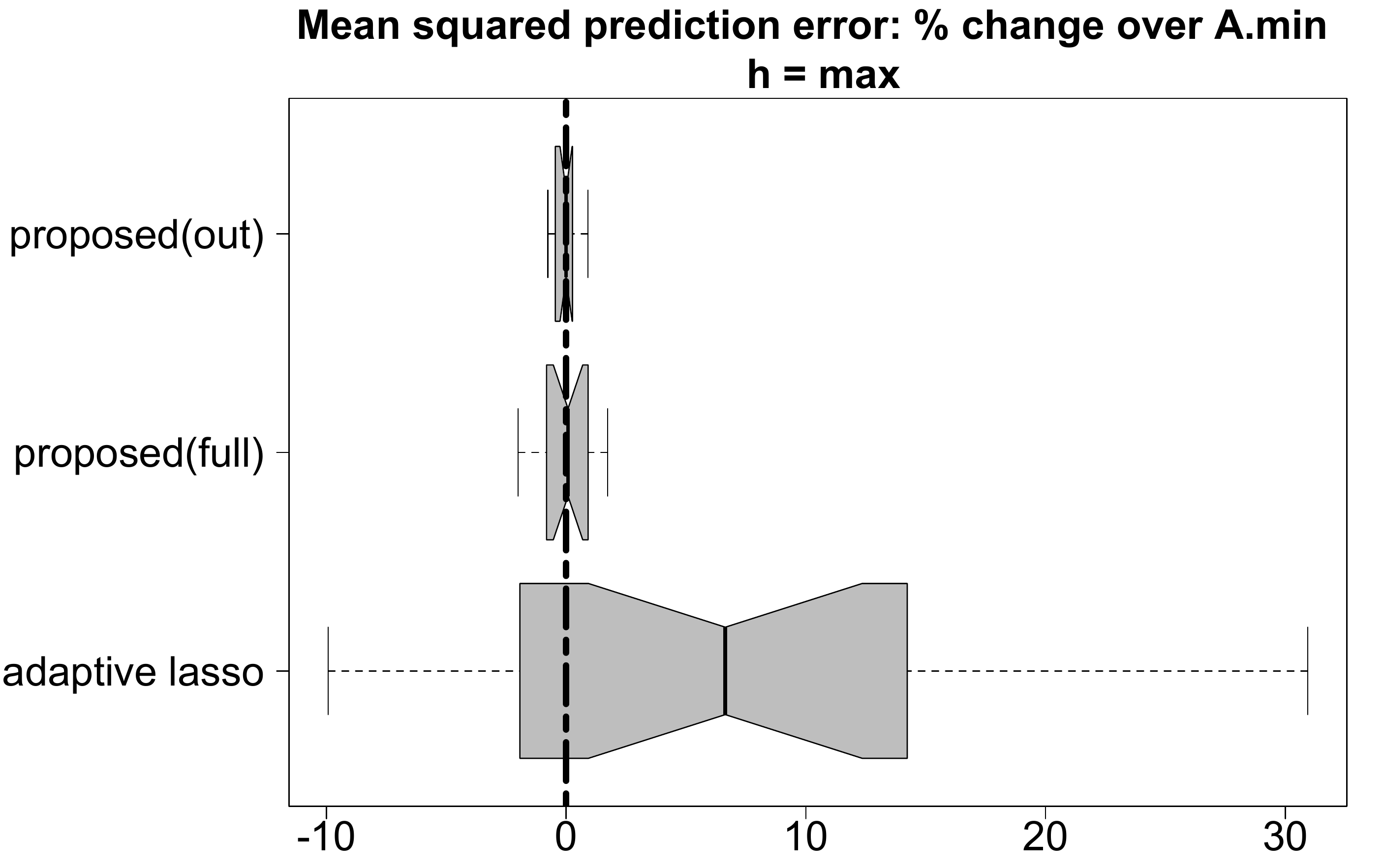}
\includegraphics[width=.32\textwidth]{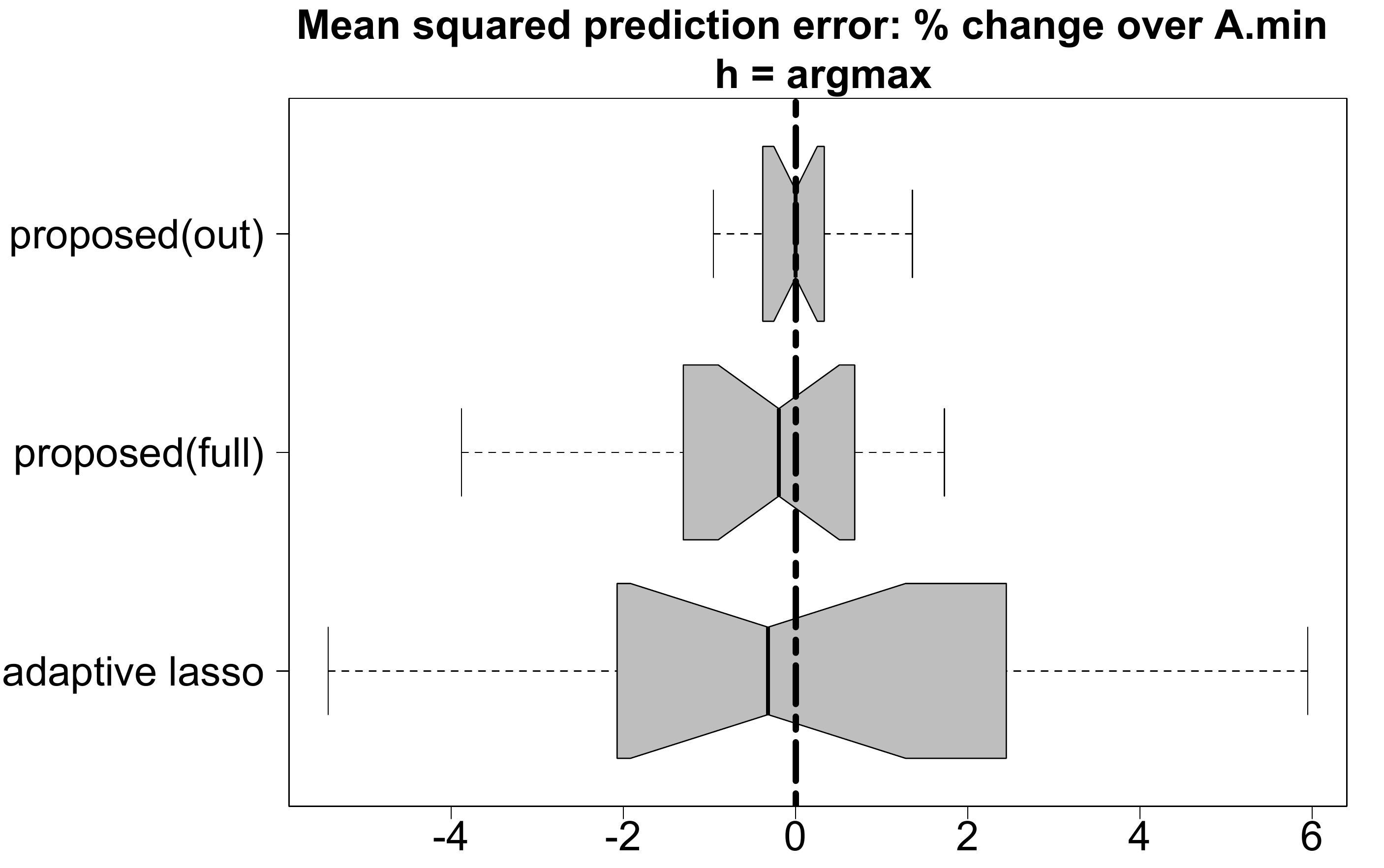}
\caption{\small  Mean squared prediction error for each functional across 20 training/validation splits. Results are presented as a percent increase relative to $\mathcal{A}_{min}$; values below zero (vertical line) indicate improvement over $\mathcal{A}_{min}$. Non-overlapping notches indicate significant differences between medians. Point predictions from $\mathcal{M}$ (\texttt{h\_bar}) are noncompetitive and omitted. Both \texttt{proposed(out)}  \texttt{proposed(full)} improve upon \texttt{adaptive lasso} and \texttt{h\_bar}, while \texttt{proposed(out)} is most accurate and performs almost identical to $\mathcal{A}_{min}$ despite using fewer covariates. \label{fig:out}}
\end{center}
\end{figure}

Point predictions were generated for the validation data using $\bar h$ under $\mathcal{M}$ (\texttt{h\_bar}), the adaptive lasso, and sparse linear actions with $\lambda= \lambda_{0, 0.1}$ (\texttt{proposed(out)}), $\lambda = 0$ (\texttt{proposed(full)}), and $\mathcal{A}_{min}$.  Since $\mathcal{A}_{min}$ is also the unique acceptable predictor when $\varepsilon=1, \eta = 0$, it provides information about robustness to $\varepsilon$ and $\eta$. The point predictions under $\mathcal{M}$ are highly inaccurate---and so excluded from Figure~\ref{fig:out}---and slow to compute: we draw $\bm{\tilde y} \sim p_{\mathcal{M}}(\bm{\tilde y} | \bm y)$ at each validation point $\bm{\tilde x}$ and then average $h(\bm{\tilde y})$ over these draws. The targeted actions simply evaluate $g(\bm{\tilde x}; \bm{\hat \delta}_{\mathcal{A}}) = \bm{\tilde x}'\bm{\hat \delta}_{\mathcal{A}}$, which is faster, simpler, less susceptible to Monte Carlo error, and empirically more accurate. Predictions were evaluated on the empirical functionals $h(\bm y_i)$ in the validation data using mean squared prediction error.

The results from the out-of-sample prediction exercise are in Figure~\ref{fig:out}.   The smallest acceptable predictor \texttt{proposed(out)} performs almost identical to the best predictor $\mathcal{A}_{min}$ despite using fewer covariates---which is precisely the goal of the acceptable predictor sets
and the out-of-sample approximations in Figure~\ref{fig:loss-out}. Both \texttt{proposed(out)} and \texttt{proposed(full)} outperform the adaptive lasso, in some cases by a large margin. The strength of this result is remarkable: the predictions are evaluated on the \emph{empirical functionals} $h(\bm y_i)$, which are used for training the adaptive lasso but \emph{not} for the proposed methods. Instead, the parametrized actions are trained using $\bar h_i$---which is itself a poor out-of-sample predictor. However, the targeted actions only rely on the in-sample adequacy of $\bar h_i$ and, unlike models trained to the empirical functionals, leverage both the model-based regularization and the uncertainty quantification provided by $\mathcal{M}$. In summary, the targeted predictors improve upon both the \emph{empirical} predictor and the \emph{model-based} predictor from which they were derived.   Lastly, we note that the performance comparisons in Figure~\ref{fig:out} confirm those in Figure~\ref{fig:loss-out}, which validates the accuracy of the out-of-sample approximations from Section~\ref{approx-oos}.

Since NHANES data are collected using a stratified multistage probability
sampling design, it is natural to question the absence of survey weights from this analysis. Although it is straightforward to incorporate the survey weights into an aggregate loss function to mimic a design-based approach (e.g., \citealp{Rao2011}), the unweighted approach has its merits. By design, NHANES oversamples certain subpopulations to ensure representation in the dataset. So although our out-of-sample predictions are not evaluated on a \emph{representative} sample of the U.S. population, they are evaluated on a \emph{carefully-curated} sample that includes key demographic, income, and age groups within the U.S. population. 



\section{Discussion}\label{discussion}
Using predictive decision analysis, we constructed optimal, simple, and efficient predictions from Bayesian models. These predictions were targeted to specific functionals and provide new avenues for model summarization. 
Out-of-sample predictive evaluations were computed using fast approximation algorithms and accompanied by predictive uncertainty quantification. Simulation studies demonstrated the prediction, estimation, and model selection capabilities of the proposed approach. The methods were applied to a large physical activity dataset, for which we built a count-valued functional regression model. Using targeted prediction with sparse linear actions, we identified 10 covariates that provide near-optimal out-of-sample predictions for important and descriptive PA functionals, with substantial gains in accuracy over both Bayesian and non-Bayesian predictors.


A core attribute of the proposed approach is that only a single Bayesian model $\mathcal{M}$ is required. The model $\mathcal{M}$ is used to construct, evaluate, and compare among targeted predictors for each functional $h$, and is the vessel for all subsequent uncertainty quantification. Although it is practically impossible for $\mathcal{M}$ to be adequate for every functional, many well-designed models are capable of describing multiple functionals. 
We only require that $\mathcal{M}$ provides a sufficiently accurate predictive distribution for each $h(\bm{\tilde y})$, which is empirically verifiable through standard posterior predictive diagnostics \citep{gelman1996posterior}. When the predictive distribution of $\mathcal{M}$ is intractable or computationally prohibitive, the proposed methods remain compatible with any approximation algorithm for $p_{\mathcal{M}}\{h(\bm{\tilde y}) | \bm y\}$. 

Future work will establish uncertainty quantification for the optimal point prediction parameters $\bm{\hat \delta}_{\mathcal{A}}$. This task is nontrivial: frequentist uncertainty estimates for penalized regression are generally \emph{not} valid, since the data have already been used to obtain the posterior (predictive) distribution under model $\mathcal{M}$. A promising alternative is to project the predictive targets $h(\bm{\tilde y})$ onto $g(\bm{\tilde x}; \bm \delta)$, which induces a predictive distribution for the resulting parameter $\bm \delta$. Similar posterior projections have proven useful for linear variable selection \citep{woody2019model} with growing theoretical justification \citep{patra2018constrained}.

\singlespacing

\ifblinded
\else 
\subsection*{Acknowledgements} 
Research was sponsored by the Army Research Office (W911NF-20-1-0184) and the National Institute of Environmental Health Sciences of the National Institutes of Health (R01ES028819). The content, views, and conclusions  contained in this document are those of the authors and should not be interpreted as representing the official policies, either expressed or implied, of the Army Research Office, the National Institutes of Health, or the U.S. Government. The U.S. Government is authorized to reproduce and distribute reprints for Government purposes notwithstanding any copyright notation herein. 
\fi


\bibliographystyle{apalike}
\bibliography{refs.bib}

\end{document}